\documentclass[]{spie}  

\usepackage{amsmath,amsfonts,amssymb}
\usepackage{graphicx}
\usepackage[colorlinks=true, allcolors=blue]{hyperref}

\usepackage{booktabs}
\usepackage{comment}

\newcommand{\TEQUILA}{{\tt TEQUILA}}
\newcommand{\Polarsens}{Polarsens\texttrademark}

\newlength{\wboxwidth}
\newcommand{\wbox}[2]{%
    \settowidth{\wboxwidth}{#1}%
    \makebox[\wboxwidth][l]{#2}%
}

\newcommand{\us}{u_\mathrm{s}}
\newcommand{\qs}{q_\mathrm{s}}
\newcommand{\pt}{p_\mathrm{t}}

\newcommand{\sbar}{\overline{s}}
\newcommand{\mbar}{\overline{m}}

\newcommand{\mathand}{\quad\mbox{and}\\}

\renewcommand{\deg}{\ensuremath{^\circ}}

\title{TEQUILA: Mechanism-free polarimetry for astronomy}

\author[a]{Alan M. Watson}
\author[b]{Noémie Globus}
\affil[a]{Instituto de Astronom\'ia, Universidad Nacional Aut\'onoma de M\'exico,\authorcr Ciudad de M\'exico, M\'exico}
\affil[b]{Instituto de Astronom\'ia, Universidad Nacional Aut\'onoma de M\'exico,\authorcr Ensenada, Baja California, M\'exico}

\authorinfo{Further author information: (Send correspondence to N.G.)\\N.G.: E-mail: globus@astro.unam.mx\\  A.M.W.: E-mail: alan@astro.unam.mx}

\begin{document} 
\maketitle

\begin{abstract}
{\TEQUILA} (Transient Event $Q$, $U$, and $I$ Light Analyzer) is an optical imaging polarimeter developed for the second Nasmyth port of the 1.3-m COLIBR\'I altitude-azimuth telescope at Observatorio Astronómico Nacional in San Pedro Mártir, México (OAN-SPM). {\TEQUILA} uses a CMOS sensor with an on-chip wire-grid micro-polarizer array to obtain simultaneous, single-exposure measurements of the Stokes parameters $I$, $Q$, and $U$ without moving optical components. This mechanism-free instrument, built entirely from commercial components, delivers seeing-limited imaging in a fixed optical band and is optimized for early-time follow-up of transient sources, including gamma-ray burst afterglows, blazars, and variable young stellar objects. In this paper, we describe the scientific motivation, the instrument design and implementation, the calibration, and initial science results.
Sensor characterization reveals a polarimetric structure in the flat field and a low quantum efficiency, which we estimate to be approximately 17\%, including losses introduced by the micro-polarizer array.
For point sources, {\TEQUILA} achieves absolute polarimetry with RMS uncertainties of 0.15\% in pupil-tracking observations and 0.20\% in field-tracking observations. 
In pupil-tracking mode, the observed RMS is fully explained by the measurement and standard-star uncertainties, with no evidence for an additional calibration term. In contrast, field-tracking observations require an additional calibration uncertainty of approximately 0.10\%. Calibration for resolved-source polarimetry remains in progress.
\end{abstract}

\keywords{polarimetry}

\section{INTRODUCTION}
\label{sec:intro}

Perhaps one of the earliest optical polarimetric observations dates back to 1819, when the physicist and astronomer François Arago made a remarkable discovery while observing the Great Comet C/1819 N1\cite{Arago1858}. Arago noticed that the light from the comet’s tail behaved differently from ordinary starlight.  Building on Malus' work \cite{Malus1810}, who had shown that reflected light becomes polarized, Arago realized that the comet's tail was reflecting sunlight scattered by dust. This was one of the first demonstrations that polarization can reveal physical information inaccessible through brightness alone.

Two centuries later, modern astronomy in the 2020s is dominated by time-domain, data-driven, and automated observing, enabling a golden age in the study of relativistic outflows. Because optical observations can be obtained with rapid cadence using robotic telescopes, optical polarimetry is particularly well suited for probing the time-dependent geometry and magnetic-field structure of astronomical transients. In these sources, polarization can vary on timescales of minutes to hours and is often strongest at the earliest times. Optical polarimetry is especially sensitive to scattering processes, synchrotron emission, and dust-induced polarization, making it highly effective for studying supernova asymmetries, circumstellar disks, blazar variability, and the early afterglows of gamma-ray bursts.

Polarization is a direct observational signature of broken symmetry. A perfectly symmetric unresolved source generally produces zero net polarization because contributions from different regions cancel. Net polarization appears when that symmetry is broken geometrically, magnetically, kinematically, or through propagation effects. Several effects produce or modify line polarization by introducing {\it a preferred direction into atomic radiation processes.} In the Zeeman effect, a magnetic field splits atomic energy levels and creates polarized spectral lines. The Stark effect does something similar using an electric field instead of a magnetic field. In the Hanle effect, magnetic fields alter the quantum interactions between atomic states, affecting the polarization of emitted or scattered lines. Continuum polarization arises from processes that introduce {\it a preferred direction into the radiation field.}  
Polarization from aligned dust grains arises because elongated grains acquire a common orientation axis, usually imposed by magnetic fields. Synchrotron radiation reflects the presence of an ordered magnetic field direction that organizes the motion of relativistic electrons. Additional propagation effects such as Faraday rotation and vacuum birefringence near strongly magnetized neutron stars further modify polarization by causing different polarization modes to propagate differently through magnetized plasma or the quantum vacuum itself.

Of the mechanisms discussed above, continuum polarization processes are especially important in the study of relativistic outflows and astronomical transients — such as gamma-ray bursts, supernovae, tidal disruption events, pulsar wind nebulae, and active galactic nuclei — where polarimetry provides unique information about magnetic-field geometry, jet structure, particle acceleration, shock physics, and the asymmetry and evolution of the emitting region. Magnetic fields are expected to play a central role in relativistic jet launching, but their topology and evolution remain only partially constrained by observations. Optical polarimetry provides one of the few direct probes of magnetic fields in these jets, with the linear polarization fraction tracing field coherence and the polarization angle reflecting the projected field structure. Comparing polarization degree and angle across wavelengths provides a tomographic view of relativistic outflows: correlated polarization behavior can reveal whether the emission originates from the same physical region, while differences between radio, optical, and X-ray polarization constrain magnetic-field stratification, jet composition, and angular jet structure. 

Optical polarimetric observations during the first minutes of the transient evolution — when polarization is expected to be strongest and most diagnostic — remain rare. 
Instruments such as {\tt FORS2} \cite{FORS2}, {\tt HOWPol} \cite{HOWPol}, {\tt RoboPol} \cite{RoboPol}, {\tt YFPOL} \cite{YFPOL}, {\tt RINGO3}\cite{RINGO3} and its successor {\tt MOPTOP}\cite{MOPTOP}, and {\tt DIPOL-1} \cite{DIPOL1} have produced important polarimetric results for relativistic jet sources.  While these instruments have revealed significant polarization variability in blazars and early-time signatures in gamma-ray burst afterglows, their limited cadence and response times have prevented continuous tracking of the earliest and most diagnostic phases of jet evolution. As a result, the connection between jet launching, magnetic-field structure, and subsequent turbulent dissipation remains incompletely understood.

The Franco-Mexican 1.3-m COLIBRÍ telescope at OAN-SPM was designed for autonomous follow-up of transient events,  principally gamma-ray bursts, in the optical and infrared with response times of 30 seconds or less \cite{Basa2022,Basa2026}. 
Its polarimetric instrument, {\TEQUILA} (Transient Events $Q$, $U$, and $I$ Light Analyzer), presented here for the first time to the community, enables high-cadence polarimetric observations by measuring the Stokes parameters $I$, $Q$, and $U$ simultaneously in a single exposure and without moving optical components. This design minimizes systematic errors associated with temporal variability and mechanical modulation while maximizing observing efficiency during rapidly evolving events. The combination of COLIBRÍ and TEQUILA  enables early-time polarimetric monitoring of relativistic transients and opens a largely unexplored observational window into the physics of relativistic jets and other transients. 

The paper is organized as follows. We first present {\TEQUILA} in section~\ref{sec:tequila}. We then discuss the electronic and cooling performance of the sensor in section~\ref{sec:sensor-characterization}. In sections~\ref{sec:laboratory-calibration} and \ref{sec:on-sky-calibration} we discuss the polarimetric calibration of the instrument first in the laboratory and then on sky. In section~\ref{sec:flat-fields} we characterize the polarimetric structure in the instrument flat field. In section \ref{sec:remaining-calibration-issues} we discuss some remaining calibration issues. We present an early example science result in section~\ref{sec:first-science}. Finally, in section~\ref{sec:pontification}, we discuss our experience and comment on future applications of these sensors. 

\section{TEQUILA}\label{sec:tequila}

The {\TEQUILA} imaging polarimeter is mounted at a Nasmyth focus of the 1.3 m COLIBRÍ robotic telescope.
The other focus mounts the DDRAGO two-channel, optical, wide-field camera\cite{2024SPIE13096E..3DL}, and in the near future the CAGIRE infrared channel will be installed. Switching between the two foci takes about 25 seconds. 

The instrument concept of {\TEQUILA} is driven by the requirement to obtain early-time optical polarization measurements of rapidly evolving sources with minimal overheads and controlled systematics.
Its optical train consists only of a fixed filter and the sensor package; the absence of moving parts enhances  stability and reliability for robotic operation. 
{\TEQUILA} had its first light on 14 November 2025.

\subsection{Wire-Grid Micro-Polarizer Sensor}
\label{sec:sensor}  

The core of {\TEQUILA} is a Sony IMX253MZR polarization-sensitive CMOS sensor.  The Sony {\Polarsens} technology was first introduced in 2016, when Sony presented its on-chip polarization imaging sensor concept \cite{Yamazaki2016}. The architecture combines a CMOS imaging sensor with a pixel-level wire-grid micro-polarizer array, enabling snapshot polarization imaging without moving optical components.  
The underlying concept of focal-plane polarization imaging using pixelated micro-polarizer arrays predates Sony {\Polarsens} by several decades and was explored in earlier imaging polarimeter research \cite{Nordin1999}. Commercial polarization sensors based on this architecture, including the IMX250MZR/MYR series, became available in 2018. Scientific publications specifically employing Sony {\Polarsens} sensors began appearing shortly after the first commercial releases. 

\begin{figure} [ht]
   \begin{center}
   \begin{tabular}{c} 
   \includegraphics[height=4.7cm]{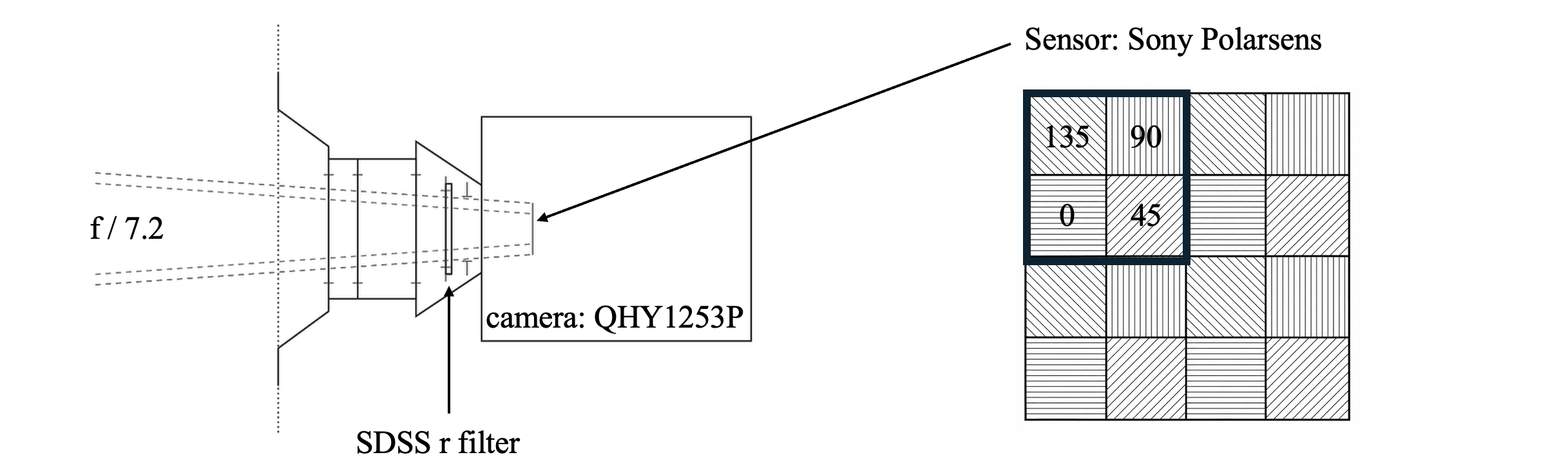}
   \end{tabular}
   \end{center}
   \caption[tequila] 
   { \label{fig:tequila} 
Illustration of the sensor architecture and sensor design.}
   \end{figure} 

The sensor employs a pixel-level micro-polarizer array, in which each pixel is covered by a linear polarizer with a specific orientation. Groups of $2\times2$ pixels form a polarimetric super-pixel that samples the incident radiation at four polarization angles, nominally $0^\circ$, $45^\circ$, $90^\circ$, and $135^\circ$, as illustrated in Figure~\ref{fig:tequila} (a). This configuration enables simultaneous reconstruction of the linear Stokes parameters $(I,Q,U)$ from each super-pixel without temporal modulation. The sensor is insensitive to circular polarization ($V$), since the micro-polarizer array contains only linear analyzers. 

Specifically, if we assume uniform illumination across a super-pixel and denote the signals in electrons from the $0\deg$, $45\deg$, $90\deg$, and $135\deg$ pixels in a super-pixel as $n_{00}$, $n_{01}$, $n_{10}$, and $n_{11}$, according to their offsets, and then nominally the linear Stokes' parameters of the illuminating light are given by
\begin{align}
I &\propto n_{00} + n_{01} + n_{10} + n_{11},\\
q \equiv Q/I &= \frac{n_{00} - n_{11}}{n_{00} + n_{11}}\label{eq:ideal-q},\mathand
u \equiv U/I &= \frac{n_{01} - n_{10}}{n_{01} + n_{10}}\label{eq:ideal-u}.
\end{align}
The degree and angle of linear polarization, $p$ and $\theta$, respectively,  are given by
\begin{align}
p &= \sqrt{q^2 + u^2},\label{eq:p}\mathand
\theta &= \frac{1}{2}\,\mathrm{atan2}(u,q),\label{eq:theta}
\end{align}
in which the factor of $1/2$ reflects the $180\deg$ symmetry of linear polarization states. The angle of polarization is therefore defined modulo $180^\circ$. These quantities form the primary science observables delivered by the instrument. We will discuss the degree to which this nominal model reflects the actual sensor in section \ref{sec:laboratory-calibration} and produce new versions of equations (\ref{eq:ideal-q}) and (\ref{eq:ideal-u}) that more accurately model the sensor.

\subsection{Sensor Package}

{\TEQUILA} uses a QHY1253P sensor package \cite{qhy1253p-data-sheet}. The key feature of this package is that it can  cool its Sony IMX253MZR sensor using a two-stage thermoelectric cooler that draws at most 12~VDC 27~W; at the time we designed {\TEQUILA} this was the only commercially-available package that combined a {\Polarsens} sensor and cooling.
The mechanical interface is C-mount, and the whole package weights about 450~g.
The control interface is USB3.0 and cooling power is supplied by a separate DC power supply.

QHY supply a software development kit (SDK) that runs both on AMD64 Linux and ARM64 Linux (e.g., a Raspberry Pi). 
Our experience with the SDK documentation is that it provides plenty of \emph{examples}, but not so much \emph{explanation}, so some experimentation was necessary. 
Nevertheless, we were able to integrate the sensor package into the TCS control system\cite{Watson2012,Basa2026} in a few days.
We use the device in single-frame mode rather than streaming mode, and it takes about 1 second to read a full frame.
(The streaming mode might be useful for fast polarimetry of very bright sources.)
At the telescope, the sensor package is controlled by a fanless industrial AMD64 PC.

\subsection{Filter}

For simplicity, stability, and robustness, TEQUILA uses a fixed Sloan Digital Sky Survey  $r$ filter (Baader \#2961700R), with half-power points at approximately 558 and 689~nm,  installed in a C-to-T-mount adapter (Baader \#2958520). This filter was chosen for two reasons. First, the polarimetric performance of the IMX253MZR sensor falls beyond 700~nm. Second, the $r$ band is commonly used for transient observations, and these are one of the key science cases for TEQUILA. 
We treat the combination of the filter and sensor package as a single unit, to ensure stability of the calibration.

\subsection{Mounting}

\begin{figure}
\centering
    \includegraphics[width=0.7\linewidth]{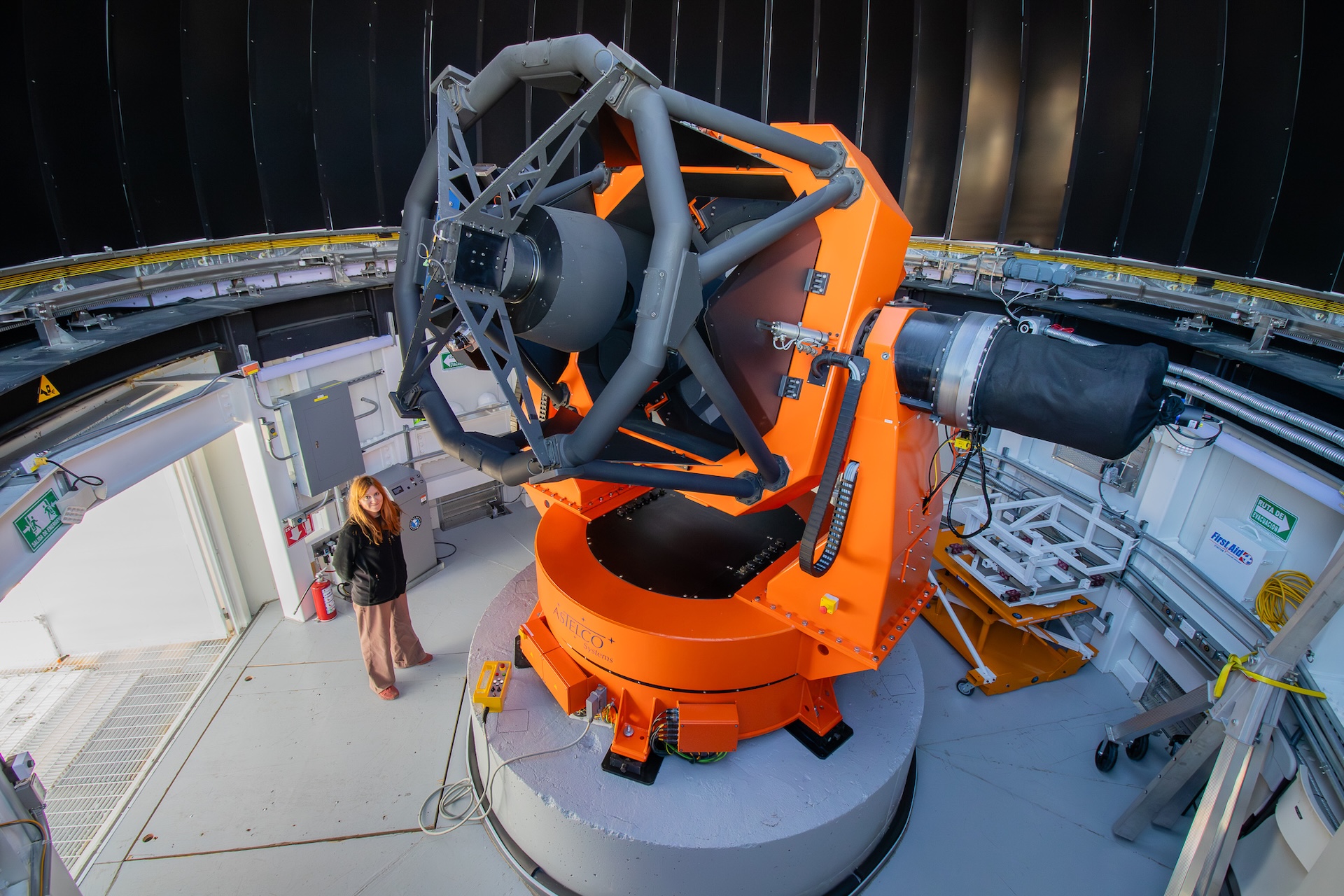}
    
    \bigskip
    
    \includegraphics[width=0.7\linewidth]{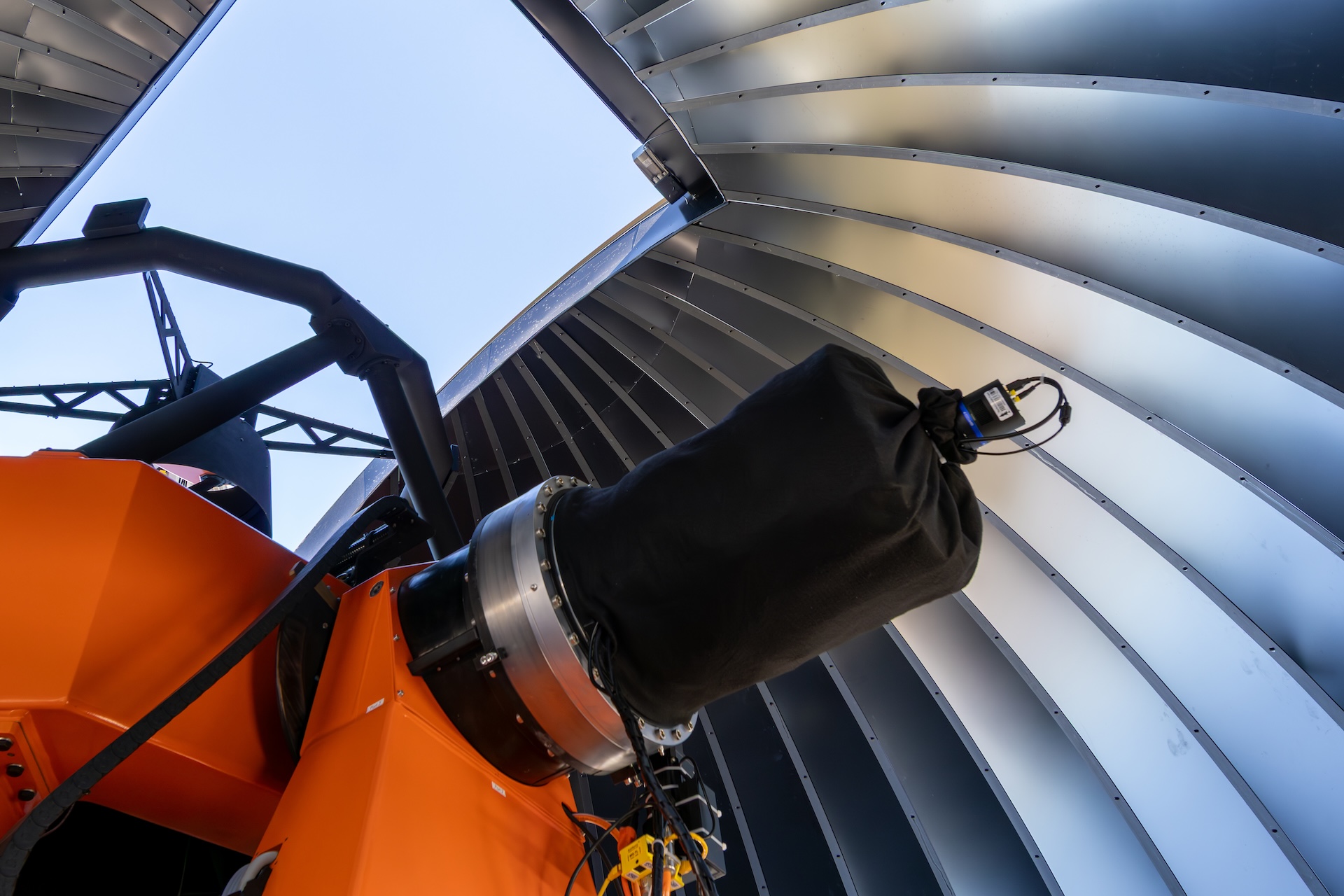}
    
    \bigskip
    
    \caption{{\TEQUILA} mounted on one of the Nasmyth foci of the COLIBRÍ telescope. {\TEQUILA} is the small black-and-blue cylinder at the right seen mounted on the larger felt-covered OGSE test-camera support structure that extends from the derotator. The purpose of this structure is simply to support an instrument close to the telescope focal plane. Note the absence of a proper cable chain and the cables simply looped from the instrument to the telescope; this forces us to limit the instrument rotation during field-tracking observations.
    The {\TEQUILA} PI is shown for scale.}
    \label{fig:colibri}
\end{figure}

Figure \ref{fig:colibri} shows the instrument on the telescope. Standard T-mount extension tubes couple the filter T-mount flange to a Celestron T-mount-to-RASA adapter on the flange of the OGSE test-camera mechanical adapter at a Nasmyth focus of the telescope and place the sensor close to the nominal focal plane. (The RASA adapter simply provides a convenient bolt pattern.)
The approximately $f/7.2$ beam from the telescope\cite{Fuentes2020} gives a scale of about 0.076 arcsec per pixel or 0.152 arcsec per super-pixel and a field of $3.8 \times 5.2$ arcmin. The fine scale ensures that even our best images with FWHMs of around 0.5 arcsec are well sampled both by pixels and super-pixels. The disadvantage is that the field is relatively small. The sensor is not vignetted.

\section{Sensor Characterization}
\label{sec:sensor-characterization}

We do not have access to detailed technical information for the Sony IMX253MZR sensor, and therefore needed to carefully characterize it as we would any other sensor. We defer discussion of its polarimetric calibration to sections \ref{sec:laboratory-calibration}, \ref{sec:on-sky-calibration}, and \ref{sec:flat-fields}.

\subsection{Format}

The SDK delivers the 12-bit DNs from the ADC\cite{sony-data-sheet} in the upper bits of 16-bit integers. We therefore shift them down by four bits to give values from 0 to 4,095 before subsequent processing.

The SDK provides images of up to $2998 \times 4128$ pixels, although smaller regions can be read. As is expected with CMOS sensors, binning in hardware is not available and would anyway defeat the polarimetric function of the sensor.
The active region appears to be the $2988 \times 4106$ pixels from columns 4 to 4109 and all rows from 0 to 2997.
(In this work, all pixel indices are zero-based.) 
The image supplied by the SDK does not seem to contain dark reference pixels. The SDK allows the pixel offset to be set and columns 4110 to 4115 uniformly contain the exact requested offset value. However, since these pixels have no noise, we suspect they are not actually read through the signal chain. Other pixels outside the active region appear to have fixed values and do not seem to contain useful information.

\subsection{Gain and Read Noise}

\begin{table}
\caption{Sensor Gain, Read Noise, and Full Well}
\label{table:gain}
\medskip
\centering
\begin{tabular}{cccc}
\toprule
Gain Setting&Read Noise&Gain&Full Well\\
&(e)&(e/DN)&(e)\\
\midrule
\phantom{00}0&3.0&2.85&\phantom{}11,600\\
\phantom{00}1&2.8&2.64&\phantom{}10,800\\
\phantom{00}2&2.8&2.67&\phantom{}10,900\\
\phantom{00}4&2.9&2.73&\phantom{}11,200\\
\phantom{00}8&2.6&2.44&\phantom{}10,000\\
\phantom{0}16&2.4&2.30&\phantom{0}9,400\\
\phantom{0}32&2.1&1.97&\phantom{0}8,100\\
\phantom{0}64&1.4&1.29&\phantom{0}5,300\\
\phantom{}128&1.3&0.63&\phantom{0}2,600\\
\phantom{}256&1.2&0.14&\phantom{00,}600\\
\bottomrule
\end{tabular}
\end{table}
 
The SDK offers different gain settings. We measured the gain and read noise for representative settings by comparing flats and short dark exposures, and show these results in Table \ref{table:gain}. As the amplifier gain increases, the read noise decreases from about 3.0 electrons to about 1.2 electrons, but this is accompanied by a dramatic increase in the gain and consequence decrease in the effective ADC saturation level from slightly more than 10,000 electrons to about 600 electrons. This behavior is similar to that shown in the data sheet\cite{qhy1253p-data-sheet}, although our measured read noise at low gain is about 2.6--3.0 electrons whereas the data sheet predicts about 1.8 electrons. The sensor also shows moderate correlated noise in the form of horizontal stripes with an RMS amplitude of about 0.33 electrons.

Since polarimetry requires high SNR, we expect to be dominated by shot noise from the source. Therefore, we have chosen to operate the device with a gain setting of 1 in order to have a higher effective full well of about 10,000 electrons, at a cost of slightly higher read noise of 2.6 electrons. The gain of 2.8 electron/DN is adequate to sample the read noise. Nevertheless, we caution that even with the lowest amplifier gain the modest effective full well limits the SNR in a single exposure to about 100 per pixel.

\subsection{Cooling Performance}

\begin{figure} [tp]
\centering
   \includegraphics[height=6cm]{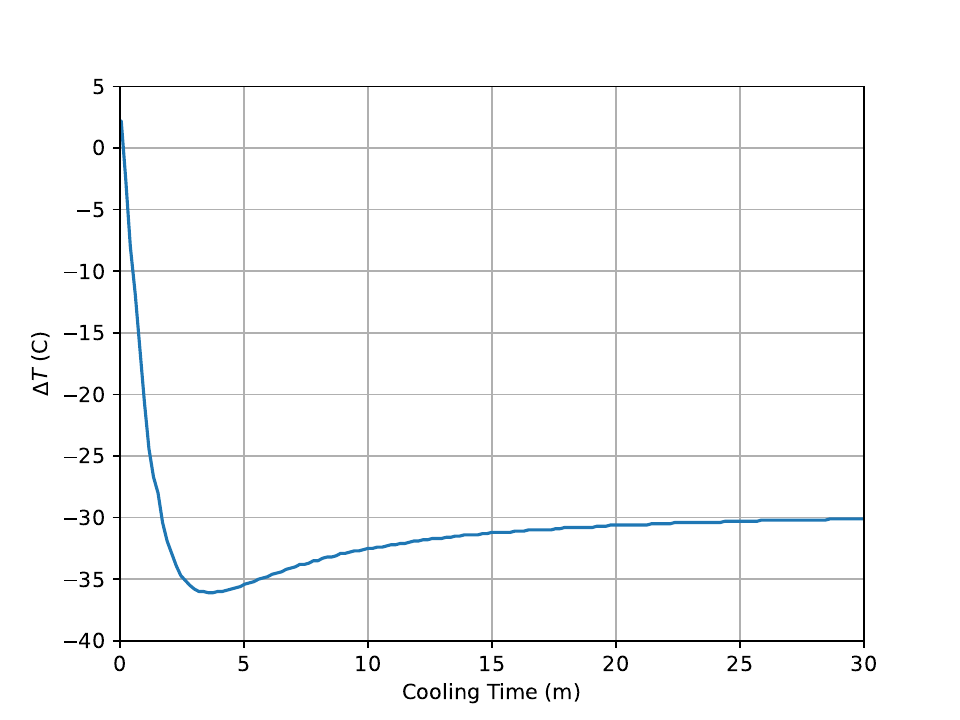}
   \caption[cooling] 
   { \label{fig:cooling} The temperature difference $\Delta T$, between the sensor and the ambient air as a function of time at full cooling power. The sensor initially cools to 35~C below ambient, but then rises to about 30~C below ambient.}
\end{figure}

The sensor package has a two-stage cooler that is specified to cool the sensor to 35~C below ambient temperature\cite{qhy1253p-data-sheet}. 
Figure \ref{fig:cooling} shows the empirical cooling performance of the sensor package at full cooling power as a function of time at about 7~C ambient temperature.
The sensor reaches 35~C below ambient in about 3 minutes, but cannot sustain this and gradually warms up to 30~C below ambient. To have a margin for stability, 25~C below ambient might be a reasonable target.
Within this range, the temperature fluctuations about the set point have a standard deviation of about 0.1 C.

The nighttime temperatures at the Observatorio Astronómico Nacional span a wide range, with 5\% and 95\% points at -4~C and +15~C\cite{2020RMxAA..56..295P}. If we wanted to use a fixed sensor temperature over all of this range, it could be no lower than about $-10$~C. 
We have been manually changing between $-20$~C and $-10$~C depending on the temperature.
We are considering automatically setting the sensor temperature at the start of each night to the lowest of $-10$, $-15$, $-20$, and $-25$~C that is no more than 25~C below the ambient temperature when the telescope opens. This should give good dark performance without requiring a large library of dark frames.

\subsection{Dark Current, Glow, and Hot Pixels}

The sensor package does not have a shutter, so we take dark images with the Nasmyth port cover closed\footnote{So, yes, we lied: {\TEQUILA} does have one mechanism.} and either at night or in twilight with the dome closed.
That said, we are unable to directly measure the dark current of the sensor due to the absence of exposed dark reference pixels and apparent systematic variations in the offset level: in long exposures the apparent signal after removing the offset is negative. Nevertheless, we have no reason to doubt the specified mean dark current of 0.02 electron/s/px or 1.2 electron/m/px at $-10$ C\cite{qhy1253p-data-sheet}. (Our preferred unit for dark current is per minute, since our exposures are typically 60 seconds.)
At this temperature, the mean dark current contribution to the noise in a 60 second exposure is only about 1 electron and so is negligible compared to the read noise. Even in a 300 second exposure, the noise contribution from the mean dark current is approximately the same as the read noise.

\begin{figure} [tp]
\centering
   \includegraphics[width=0.5\linewidth]{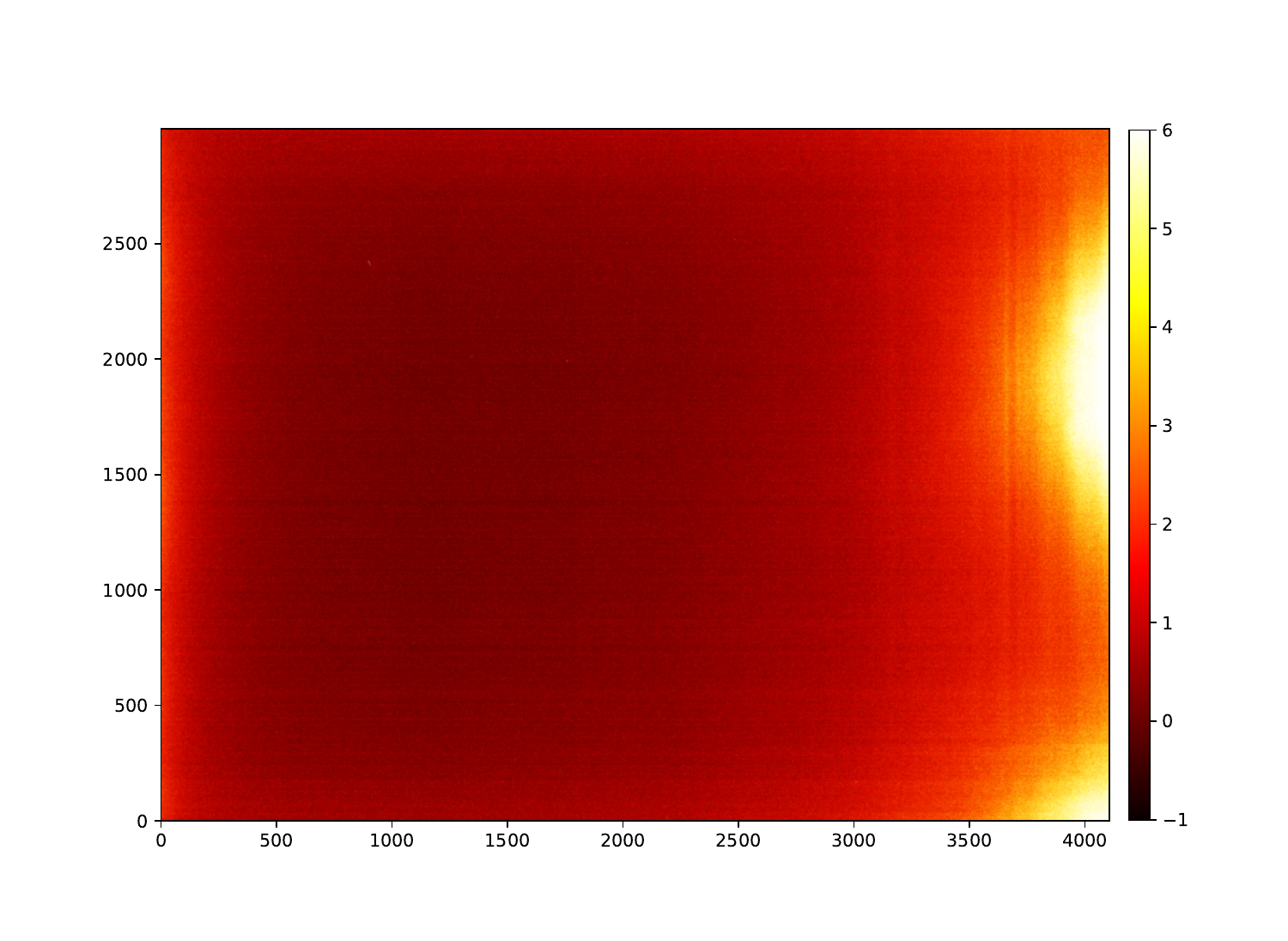}%
   \includegraphics[width=0.5\linewidth]{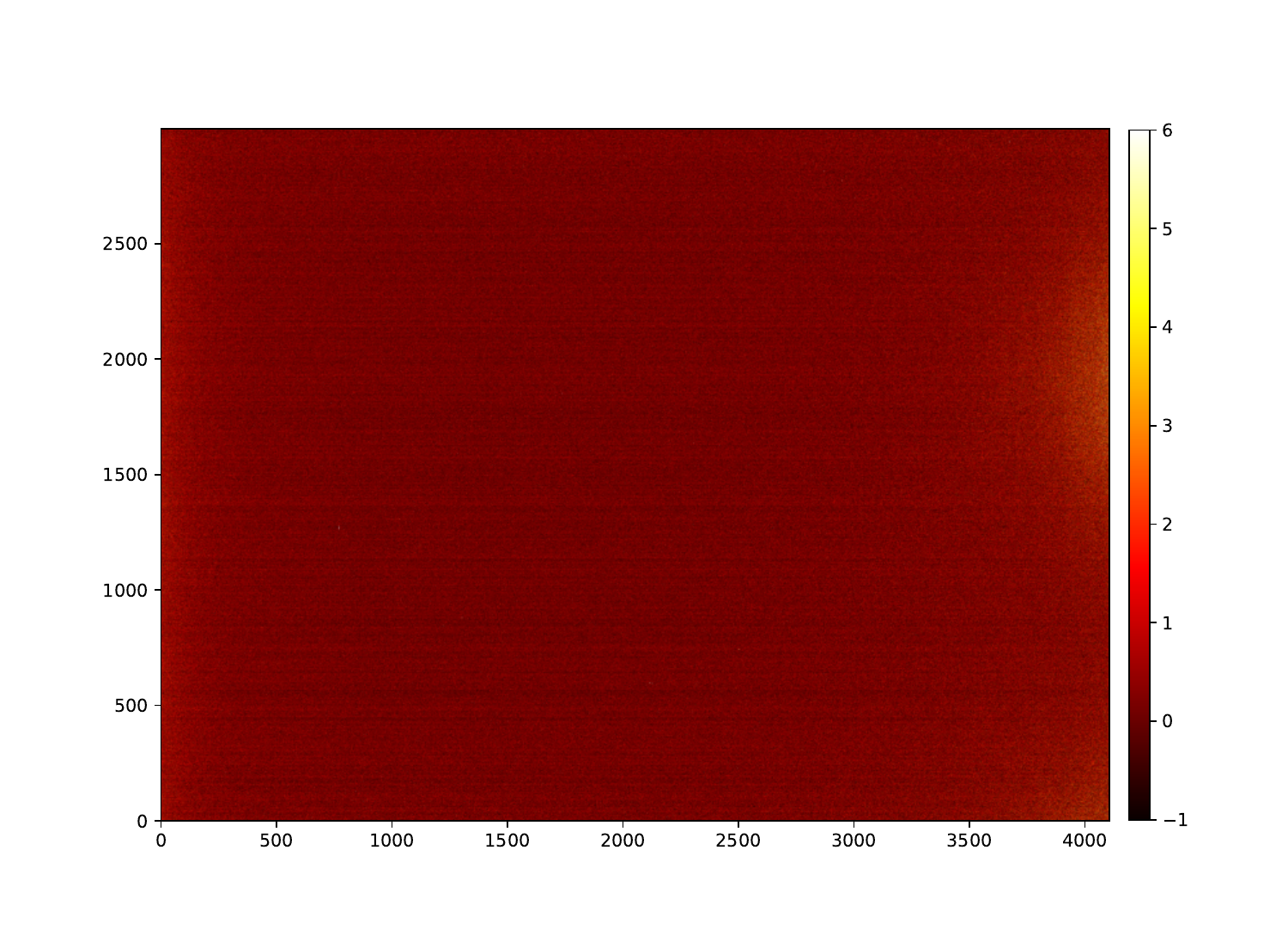}
   \caption{ \label{fig:glow} Left: The mean dark rate in electron/m/px at $-20$ C after subtracting the median. Note the bright regions on the left and right. Right: The difference in mean dark rates in electron/m/px at $-10$ C and $-20$ C, again after subtracting the median. The bright regions largely disappear, suggestion that they do not depend on temperature and are likely glow from electronics. We typically observe sources in the clear region between columns 500 and 3000.}
\end{figure}

Nevertheless, there are two additional considerations in the dark images. The first is an apparent glow that increases with exposure time but does not change with sensor temperature. Figure \ref{fig:glow} shows a 300 second dark image at $-20$ and shows bright regions on the right and to a lesser degree on the left edge. It also shows the difference between two 300 seconds dark images at $-10$ and $-20$ C, in which these regions disappear. The conclusion is that the rates in these regions do not depend on temperature and so are not the result of elevated dark current (e.g., by local heating). Instead, the simplest explanation is that they are the result of glow from the electronics in the sensor head. In these regions, the effective dark current reaches about 6 electrons/m/px, so in a one minute exposure the additional noise is approximately the same as the read noise.

\begin{figure} [tp]
\centering
   \centering
   \includegraphics[height=6.5cm]{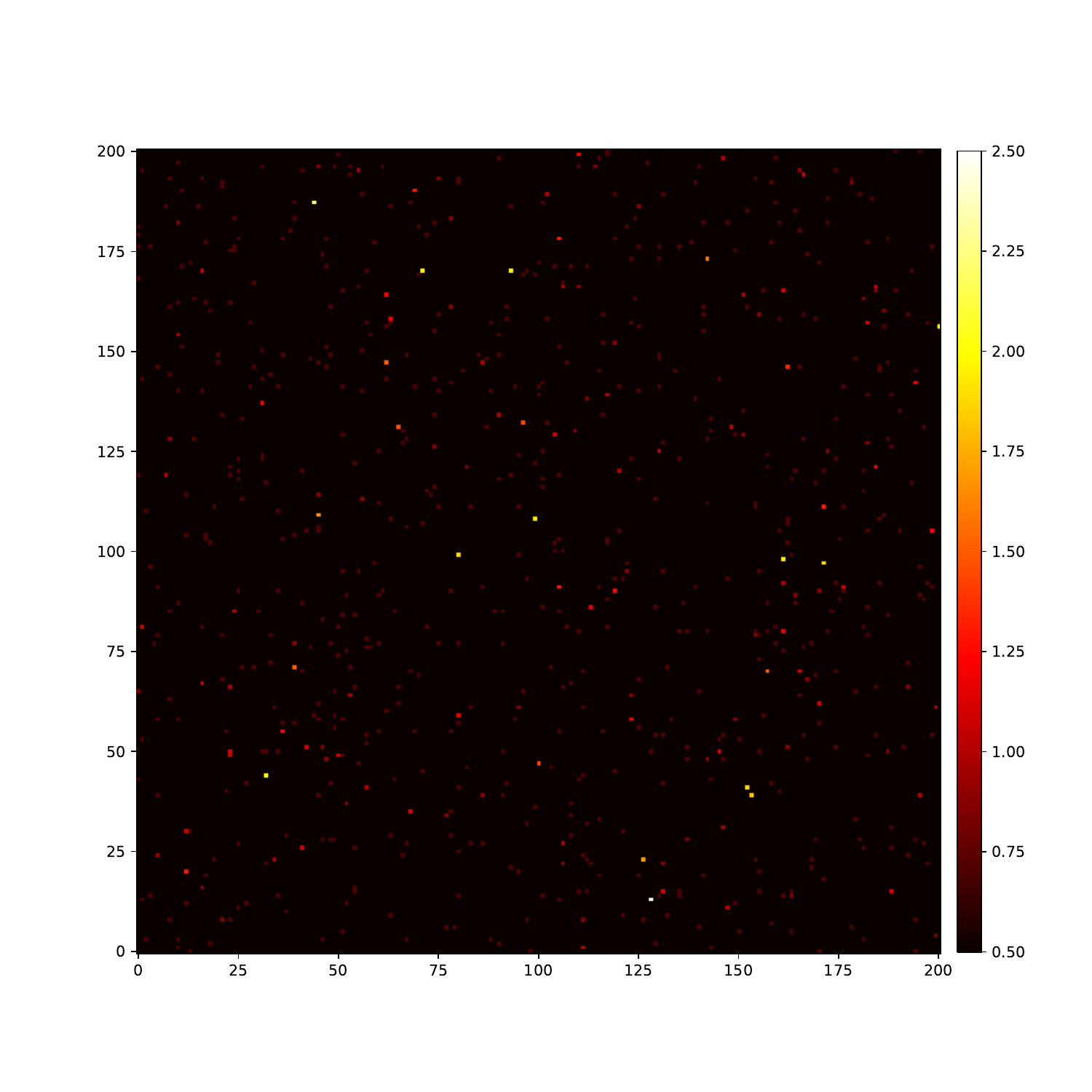}
   \raisebox{0.27cm}{%
      \includegraphics[height=6cm]{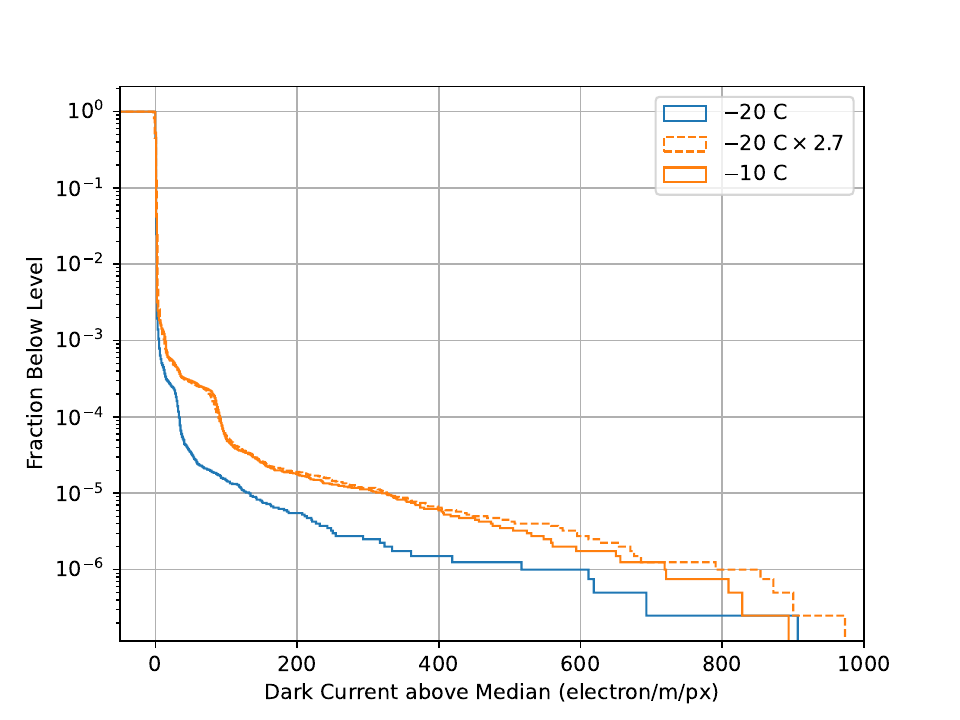}%
   }
   \caption[dark-current] 
   {Left: A representative region of a dark image at $-10$ C showing the relatively large number of hot pixels. The color map is $\log_{10}$ of the dark current in electron/m/px and ranges from approximately 3 electron/m/px (0.5 in the logarithm) to approximately 300 electron/m/px (2.5 in the logarithm). Above a rate of 10 electron/m/px (i.e., 1 in the logarithm), the noise from the dark current in a one minute exposure dominates the read noise. Right: Histograms of the dark current in electron/m/px relative to the median as a function of temperature. }
   \label{fig:dark-current}
\end{figure}

Furthermore, there is a population of hot pixels that have dark currents that significantly exceed the median rate. 
The left panel of Figure \ref{fig:dark-current} shows a representative region of the dark current image at $-10$~C with a logarithmic scale. The large number of hot pixels is not a defect of the sensor, but simply a consequence of operating it at a relatively high temperature.
Above a rate of 10 electron/m/px (i.e., 1 in the logarithm), the noise from the dark current in a one minute exposure dominates the read noise.
The right panel of Figure \ref{fig:dark-current} shows histograms of the dark current rate above the median value. Reducing the sensor temperature from $-10$ to $-20$ C reduces the dark current of these hot pixels by about a factor of 2.7, as expected from the rule of thumb that the dark current doubles every 7 C, and reduces the fraction with a rate of 10 electron/m/px or more from about 1 in 430 to about 1 in 1,200 and the fraction with a rate of 100 electron/m/px or more from about 1 in 16,000 to about 1 in 56,000.
Together, these figures show the desirability of cooling the sensor, maintaining temperature stability, and subtracting a reference dark frame at the correct temperature.

\subsection{Flat-Field Uniformity and Cosmetics}

\begin{figure} [tp]
\centering
   \includegraphics[width=0.5\linewidth]{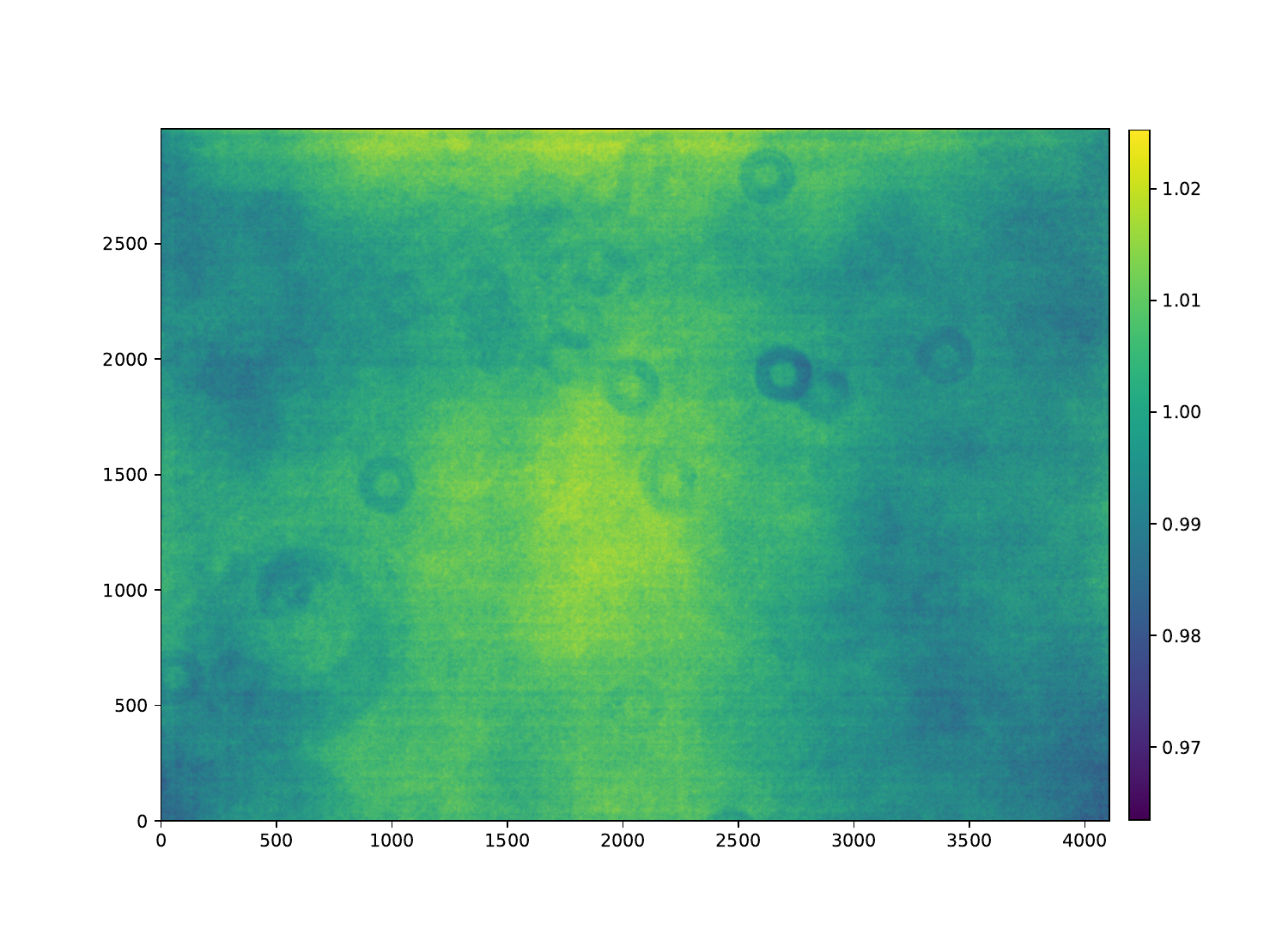}
   \caption{ \label{fig:flat} A typical mean twilight flat field image. The  pixel-to-pixel variation is about 1\% RMS and the large scale structure has an RMS of about 0.7\% and a peak-to-valley range of $\pm2\%$}
\end{figure}

The sensor shows excellent uniformity and cosmetics. In laboratory flats with unpolarized light, there are no pixels with a response of less than 80\% of the global median and none with a response of less than 90\% of the local median (in a $7 \times 7$ box). The pixel-to-pixel variation is about 1\% RMS. In twilight flats, the large scale structure has an RMS of about 0.7\% and a peak-to-valley range of $\pm2\%$.

\subsection{Quantum Efficiency}

We did not attempt to measure the quantum efficiency of the sensor in the laboratory. Instead, we estimated the effective efficiency of the TEQUILA optics and sensor on the telescope by comparing the electron count rates of unpolarized standard stars measured consecutively in the Sloan $r$ band with TEQUILA and the DDRAGO instrument \cite{2024SPIE13096E..3DL} mounted on the other Nasmyth port.
This differential comparison has the advantage that it is largely independent 
of the reflectivities of the three telescope mirrors and the atmospheric transmission.

We estimate that the system efficiency of the optics and detector in DDRAGO is 0.74 in $r$. The relative filter widths (the wavelength-integrated transmission) are 123~nm for TEQUILA, based on the vendor's transmission curve, and 134~nm for DDRAGO, based on measured transmission data. The median ratio of the electron count rates in TEQUILA and DDRAGO is 0.20, so we estimate that the quantum efficiency of the TEQUILA sensor in the Sloan $r$ band is 0.17.
The principal uncertainty in this determination comes directly from uncertainties in the gain values for the two instruments, and so is probably a relative uncertainty of about 10\%.

This value includes transmission and reflection losses in the window, sensor cover glass, sensor micro-lenses, and on-chip micro-polarizer array in addition to the intrinsic quantum efficiency of the silicon detector. A perfect linear polarizer would reduce the throughput of unpolarized light by a factor of two, but our measured efficiency is substantially below this, suggesting additional losses associated with the {\Polarsens} architecture. To our knowledge, Sony has not published a quantum-efficiency curve for the IMX253MZR, and no independent measurements have been reported in the literature. Our measured effective quantum efficiency of approximately 17\% therefore provides one of the first experimental characterizations of the sensor's sensitivity and suggests that there is potential for significant improvement in the quantum efficiency of these sensors.

\subsection{Read Problems}

We have encountered a problem with the reading the sensor. Normally, reading the sensor takes about 1 second. However, sometimes the sensor package gets into a state where it takes 60 seconds to read and returns an image of all zeros. Restarting the software seems to have no effect. In order to recover from this, we need to cycle the power to the sensor package. 

We saw this problem more frequently when running the device from a Raspberry Pi; then we would see this problem every few dozen images and encountered a range of conflicts between the sensor package and other USB devices attached to the same computer. Therefore, until this problem is fixed, we cannot recommend using this sensor package with a Raspberry Pi computer.

Since we switched to using a more powerful industrial PC with an AMD64 processor, this problem only seems to occur after two or three days. Therefore, we deal with it by rebooting the device every afternoon. While this mitigates the problem, it is not entirely satisfactory. We are therefore working with the vendor to find a robust solution. 

\section{Laboratory Polarimetric Calibration}
\label{sec:laboratory-calibration}

We calibrated the polarimetric properties of the sensor in the laboratory in order to have confidence in our on-sky measurements and to correct for some small non-ideal behavior.
In developing our calibration plan, we found especially useful discussion of polarization sensor calibration in earlier work \cite{2011SPIE.8012E..0HY,2012ApOpt..51.5392Y,2013OExpr..2121039P,2018ApOpt..57.4992F}.

\begin{figure} [pt]
    \centering
   \includegraphics[height=6.5cm]{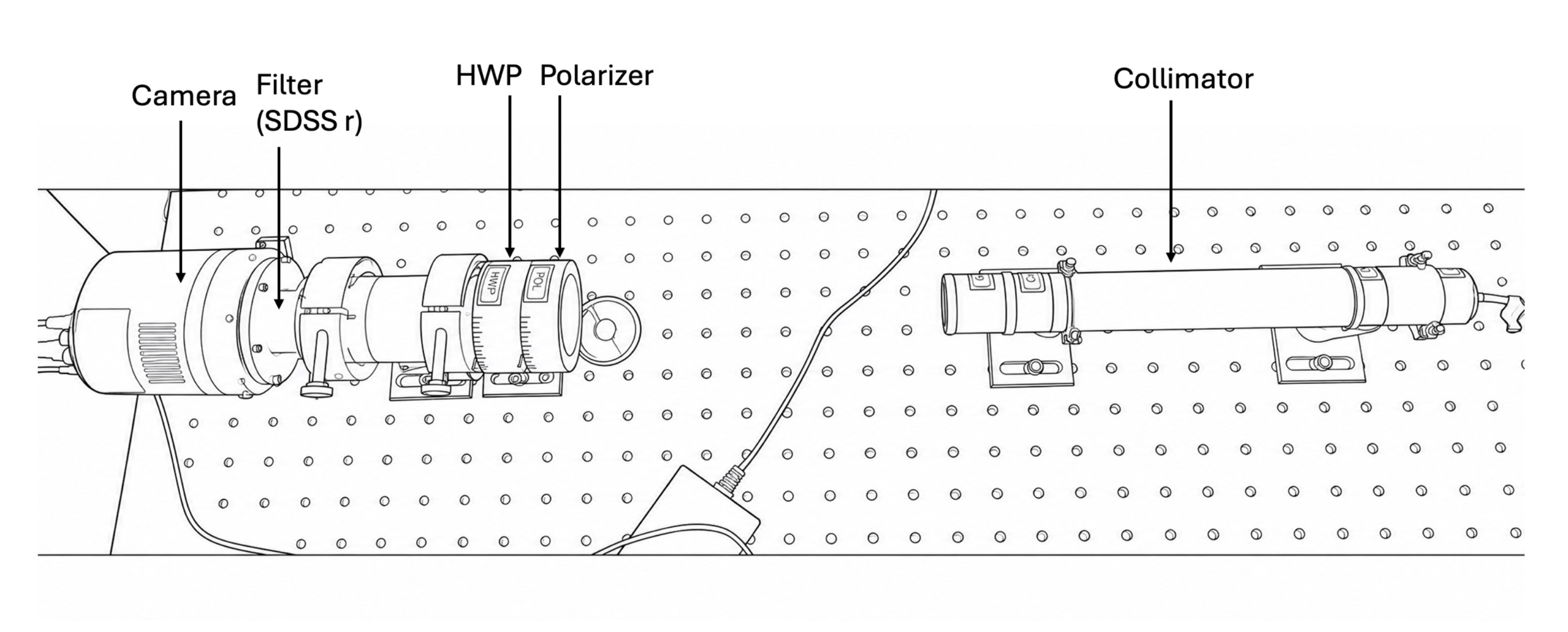}
   \caption{ \label{fig:laboratory-calibration-setup} Lab calibration set-up. The path of the collimated beam is shown in transparency when inside the optical assembly. After being collimated, the beam passes through a polarizer and a half-wave plate before hitting the sensor.}
\end{figure}

The instrumental set-up is shown in Figure~\ref{fig:laboratory-calibration-setup}. The sensor is illuminated by a quartz lamp illuminating an opal diffuser, the light from which is then collimated and directed at the sensor.
Before the sensor, a polarizer (Edmund \#19-652) generates a constant polarization that can be rotated on the sensor by a rotating half-wave plate (Edmund \#39-039). The half-wave plate and the polarizer are mounted in manual rotating mounts that couple to the T-mount flange of the instrument. For each set of images, we left the polarizer at a fixed position and rotated the half-wave plate in approximately $15\deg$ steps through slightly more than $180\deg$.

\begin{figure}[tp]
\centering
   \includegraphics[width=0.8\linewidth]{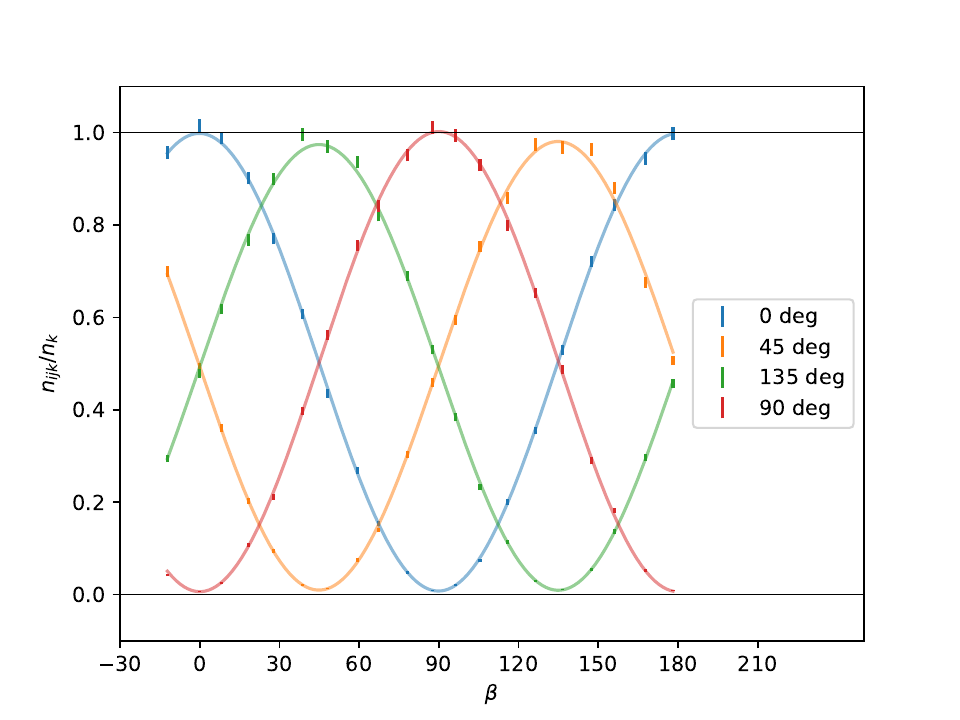}
   \caption{ \label{fig:pixel-fits} The laboratory data and corresponding model fits for pixels in a representative super-pixel. The fits have $\bar\chi^2 = 0.93$}
\end{figure}

Our model for the signal from a sensor illuminated by this source assumes the light is perfectly polarized but accounts for non-ideal polarizers, pixel-to-pixel sensitivity variations, and small orientation offsets. 
Specifically, our model is
\begin{align}
n_{xyk} = n_k \frac{s_{xy}}{1+m_{xy}} \left[\cos^2 (\alpha_{xy} + \Delta\alpha_{xy} + \beta_k) + m_{xy} \sin^2(\alpha_{xy} + \Delta\alpha_{xy} + \beta_k)\right],
\label{eq:laboratory-model}
\end{align}
in which $n_{xyk}$ is the signal in pixel $(x,y)$ in exposure $k$, $n_k$ is the mean signal in exposure $k$, $s_{xy}$ is the relative sensitivity of each pixel, $m_{xy}$ is the extinction ratio of each pixel, $\beta_k$ is the rotation of the half-wave plate in exposure $k$, $\alpha_{xy}$ is the nominal polarizer orientation ($0\deg$, $45\deg$, $90\deg$, or $135\deg$) of each pixel, and $\Delta\alpha_{xy}$ is the offset of the polarizer orientation from the nominal for each pixel.
The interpretation of $s_{xy}$ and $\alpha_{xy}$ is straightforward. Consideration of equation (\ref{eq:laboratory-model}) shows that the extinction ratio $m_{xy}$ is the relative signal that is passed when the incident polarization is perpendicular to the polarizer.

\begin{figure}[tp]
\centering
   \includegraphics[trim=2cm 0cm 1cm 0cm, clip, width=0.33\linewidth]{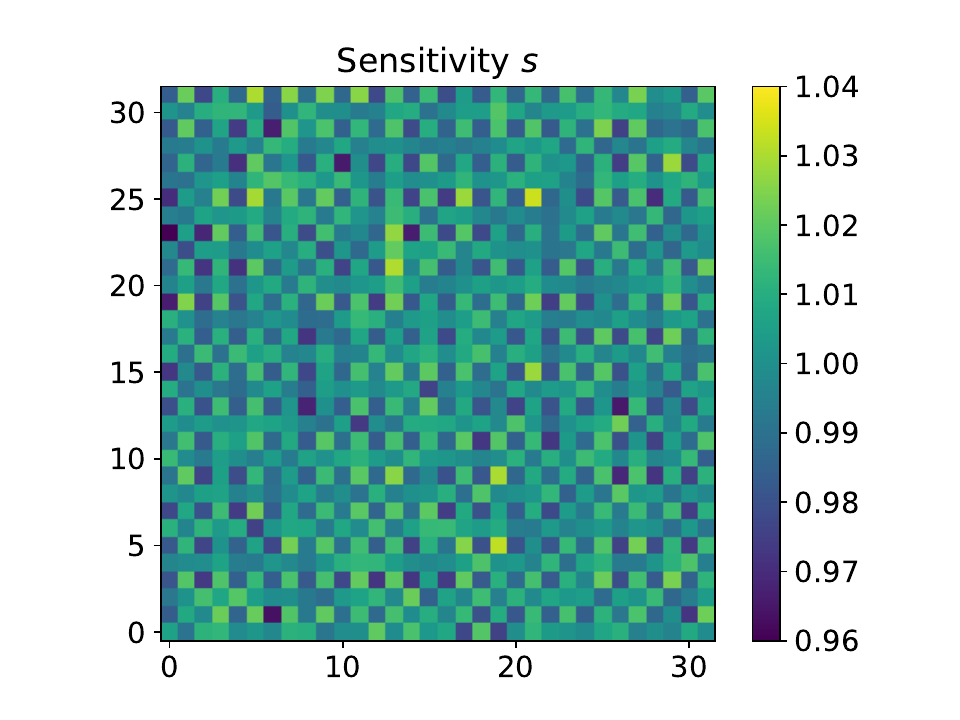}%
   \includegraphics[trim=2cm 0cm 1cm 0cm, clip, width=0.33\linewidth]{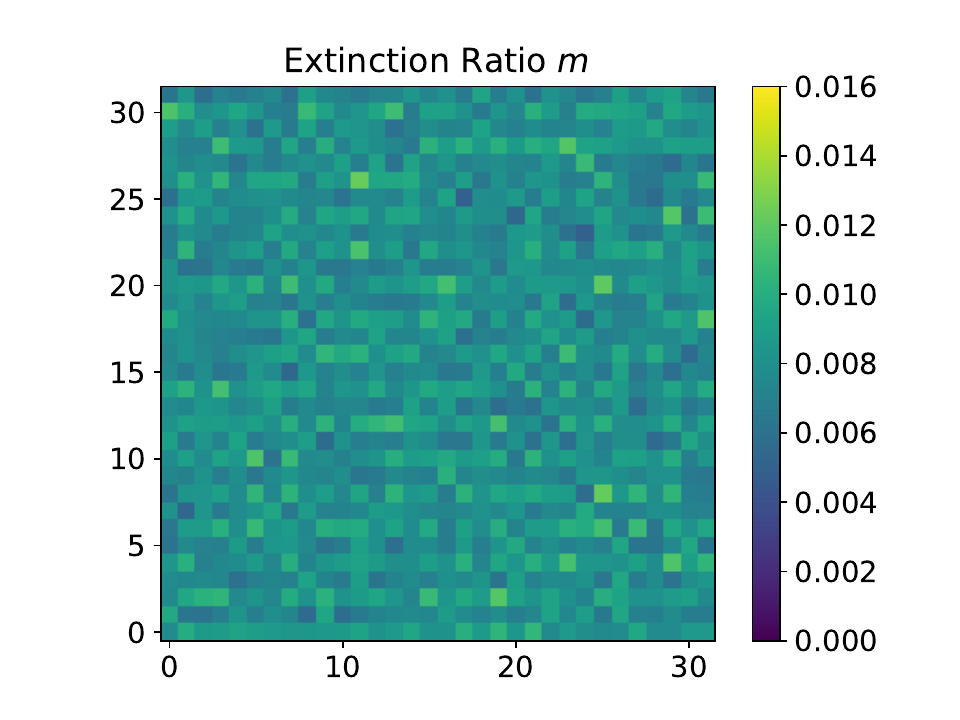}%
   \includegraphics[trim=2cm 0cm 1cm 0cm, clip, width=0.33\linewidth]{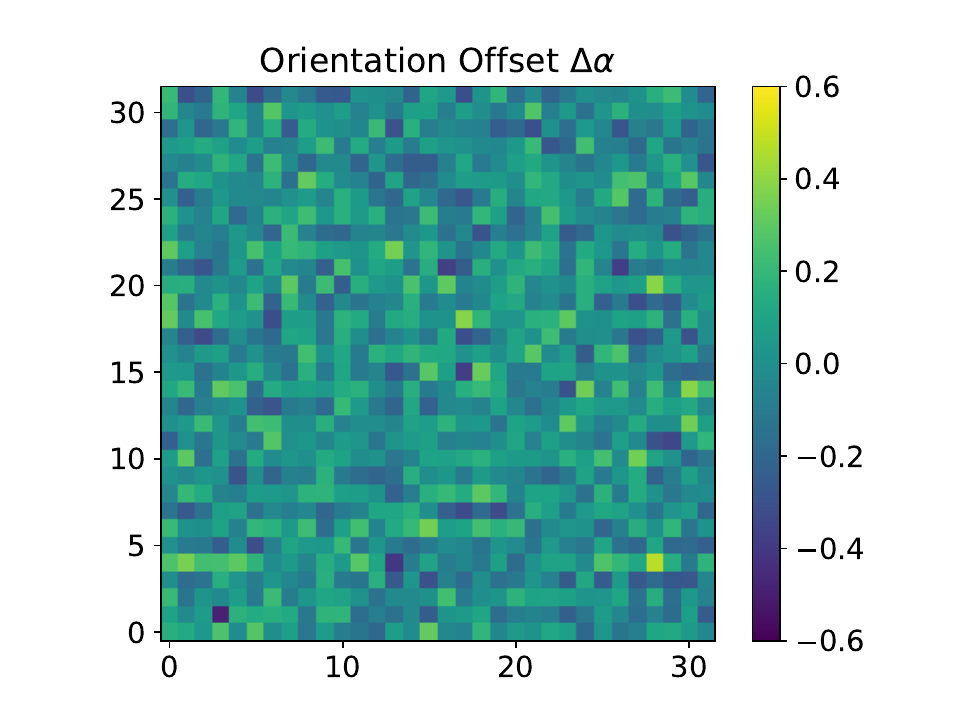}

   \caption{ \label{fig:sensor-fits} The sensor parameters $s_{xy}$, $m_{xy}$, and $\Delta\alpha_{xy}$ derived in the laboratory calibration in a representative region close to the center of the sensor.}
\end{figure}

We solved for $s_{xy}$, $m_{xy}$, and $\Delta\alpha_{xy}$ by minimizing $\chi^2$. 
Furthermore, since we cannot rotate the half-wave plate precisely, we solved iteratively for each $\beta_k$.
Figure \ref{fig:pixel-fits} shows the data values and corresponding fits for four pixels within a representative super-pixel.
Figure \ref{fig:sensor-fits} shows the values of the parameters for a representative region near the center of the sensor.
The resulting parameter distributions are found to be highly uniform across the sensor, with
\begin{align}
s &= 1.000 \pm 0.015, \\
m &= 0.007 \pm 0.001,\mathand
\Delta\alpha &= 0.00^\circ \pm 0.14^\circ.
\end{align}
The fit has $\bar\chi^2 = 1.03$, indicating that the model in equation~\eqref{eq:laboratory-model} provides an adequate description of the sensor response. We note that the polarimetric performance of the sensor is close to ideal.

Our determination of $s$ is constrained so that the average $\sbar$ is 1 and is unable to separate large-scale changes in the efficiency of the sensor from non-uniform illumination. However, this is not a real problem in practice, since our on-sky flats show that the sensor is highly uniform and because polarization parameters are determined locally. Furthermore, our determination of $\Delta\alpha$ is constrained so that the average $\overline{\Delta\alpha}$ is zero, but again this is not a problem as the polarization position angle on the sky is determined by observations of polarized standards.

We see that the variation in the extinction ratio $m$ is small, so to good precision we can treat every pixel as if it had the average extinction ratio $\bar m = 0.007$.
We can see the impact of this non-zero mean extinction ratio by noting that the signal from a real sensor ($\mbar > 0$) is equivalent to that of an ideal sensor ($m \equiv 0$) in which a fraction $\mbar/(1+\mbar) \approx \mbar$ of the linearly polarized light is converted into unpolarized light. Thus, a real sensor will underestimate the degree of linear polarization by a factor of approximately $(1-\mbar)$. This can be corrected by multiplying the derived linear polarization for a real sensor by the corresponding inverse correction factor $1/(1-\mbar)\approx(1+\mbar)$. For our sensor, this correction factor is 1.007 and in real observations is dwarfed by other uncertainties. 
Nevertheless, including it in our calibration is easy, since it is just a multiplicative factor. 

Our corrected model for the normalized Stokes' parameters $q$ and $u$ for each super-pixel, including sensitivity variations and a mean extinction ratio but ignoring variations in the micro-polarizer orientations, is therefore
\begin{align}
n'_{ij} &= n_{ij} / s_{ij},\\
q &= (1+\mbar)\left[\frac{n'_{00} - n'_{11}}{n'_{00} + n'_{11}}\right],\mathand
u &= (1+\mbar)\left[\frac{n'_{01} - n'_{10}}{n'_{01} + n'_{10}}\right].
\end{align}
These equations supersede (\ref{eq:ideal-q}) and (\ref{eq:ideal-u}).

We caution that the data sheet for the sensor\cite{sony-data-sheet} suggests that the extinction ratio depends on wavelength, with $\mbar$ being lower in the blue and higher in the red, and so needs to be calibrated for each filter in an imaging polarimeter. {\TEQUILA} only has one filter, so we only need to adopt a single value of $\mbar$.

\section{On-Sky Polarimetric Calibration}
\label{sec:on-sky-calibration}

Polarimeters on altitude-azimuth telescopes are often calibrated using an approach based on determining the Mueller matrix for each component, with appropriate rotations\cite{2003SPIE.4843..456G,2018PASA...35...12W,2019MNRAS.482.5023H}. This is especially advantageous in instruments with rotating half-wave and quarter-wave plates and those that measure circular polarization. 

The polarimetric calibration of {\TEQUILA} is simpler than these more general cases for a number of reasons. First, {\TEQUILA} can only measure intensity and linear polarization, but not circular polarization. Second, we explicitly assume that the incoming light has zero circular polarization and ignore any cross-talk between $V$ and $Q$ or $U$. Third, we expect the classical instrumental polarization will have a component from the telescope, dominated by the reflection from the tertiary mirror, that rotates with the parallactic angle and another component that rotates with the instrument. Therefore, we will adopt a geometric approach to the polarimetric calibration.

\subsection{Calibration Observations}

To calibrate the instrument, we observed low- and high-polarization RoboPol standards\cite{2023A&A...677A.144B} with magnitudes of about 10 to 14.
For each standard, we observed a $3\times3$ grid aligned N--S with a grid spacing of $15''$ and with two exposures of 10 seconds at each grid position.

We reduce each exposure by subtracting a mean dark and dividing by a mean flat. We obtain evening twilight flat field images each night with the telescope typically pointed west and at a zenith distance of $70\deg$. These images are of course polarized, both because the twilight sky is polarized and because of the reflection off the tertiary mirror. Therefore, we normalize each channel independently using the median in the central part of the sensor filtering and averaging to produce the mean flat field. The flat fields are discussed in detail in section \ref{sec:flat-fields}.

For each exposure, we form separated images by extracting the $0\deg$, $45\deg$, $90\deg$, and $135\deg$ pixels from each super-pixel. We denote these as 00, 01, 11, and 10 by their offsets within each super-pixel. We average these separated images to form an intensity image.
We perform photometry of the standards with the {\tt photutils} python package\cite{bradley_2026_19636730}. We find stars in the intensity image, as the pixel-to-pixel variations in the complete image due to polarization for example from scattered Moonlight appear to star finders as an unanticipated source of noise.
We estimate the FWHM of the stars and perform aperture photometry on the four separated images, with a star aperture of radius 1.0 FWHM and a sky annulus between radii of 2.0 and 3.0 FWHM.
We note that the pixel centers in the separated images are offset from those in the intensity image by $\pm1/4$ pixel in each direction, and so we adjust the centers of the stars appropriately for photometry of each separated image.
We estimate the background noise in the star aperture using the measured noise in the sky aperture and the source noise using photon statistics and the measured gain. 

The products of the photometry are the sensitivity- and background-corrected aperture signals $N'_{ij} \equiv \sum n_{xy}/s_{xy}$ for each star and their
associated uncertainties $\sigma_{ij}$.
We can then calculate raw polarizations $\qs$ and $\us$ as
\begin{align}
    \qs &= (1+\mbar)\left[
    \frac{N'_{00}-N'_{11}}
         {N'_{00}+N'_{11}}\right],\mathand
    \us &= (1+\mbar)\left[
    \frac{N'_{01}-N'_{10}}
         {N'_{01}+N'_{10}}\right],
\end{align}
in which $\mbar=0.007$ is the mean extinction ratio determined in the
laboratory. The corresponding uncertainties are
\begin{align}
    \sigma(\qs) &=
    \frac{2(1+\mbar)}{(N'_{00}+N'_{11})^2}
    \left[
        (\sigma_{11} N'_{00})^2
        +
        (\sigma_{00} N'_{11})^2
    \right]^{1/2},\mathand
    \sigma(\us) &=
    \frac{2(1+\mbar)}{(N'_{01}+N'_{10})^2}
    \left[
        (\sigma_{10} N'_{01})^2
        +
        (\sigma_{01} N'_{10})^2
    \right]^{1/2}.
\end{align}

\subsection{Sensitivity}\label{ssec:sensitivity}

\begin{figure}
\centering
    \includegraphics[width=0.65\linewidth]{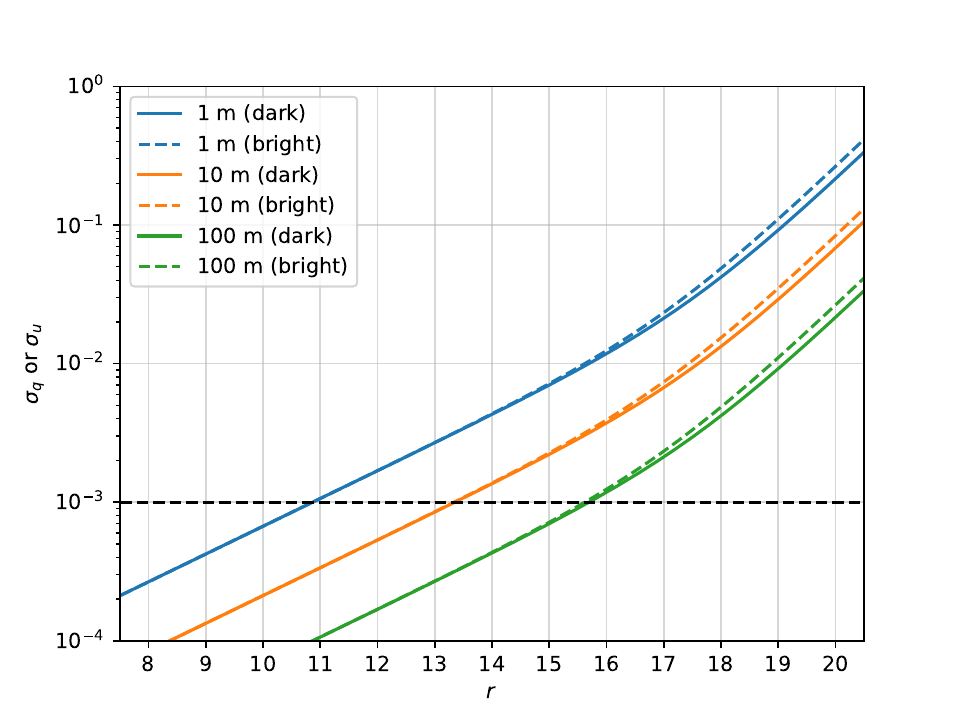}

    \caption{The current estimated polarimetric uncertainty $\sigma_q$ and $\sigma_u$ as a function of $r$ magnitude for exposure times of 1, 10, and 100 minutes in bright and dark time. The model is calibrated against observations. The dashed black line shows 0.1\%, the current approximate limit of our demonstrated precision. We see that in 1 minute we expect 1\% uncertainty at $r \approx 15.5$ and in 10 minutes we expect 0.1\% uncertainty at $r\approx 13.5$. We expect the sensitivity to improve after recoating the mirrors in September 2026.
    }
    \label{fig:sensitivity}
\end{figure}

For a star of magnitude 10, TEQUILA on COLIBRÍ gives a total signal of approximately $10^{4.9}$ electron/s. For a star with low polarization, the uncertainty in $q$ and $u$ is related to the uncertainty $\sigma_I$ in the total intensity $I$ by $\sigma_q \approx \sigma_u \approx \sqrt{2}\sigma_I/I$.  For sufficiently bright stars, where photon noise dominates, $I/\sigma_I \approx \sqrt{N}$, in which $N$ is the total signal in electrons. Thus, we can estimate the polarimetric sensitivity for an exposure of $t$ seconds on a point source of a given $r$ magnitude as
\begin{equation}
\sigma_q = \sigma_u \approx 5 \times 10^{-5}t^{-1/2} 10^{0.2r} .
\end{equation}
A more sophisticated model, including read noise, sky background, and limiting individual exposures to 60 seconds, is shown graphically in Figure~\ref{fig:sensitivity}. We see that in 1 minute we expect 1\% uncertainty at $r \approx 15.5$ and in 10 minutes we expect 0.1\% uncertainty at $r\approx 13.5$.

The current sensitivity is limited in part by the degraded reflectivity of the COLIBRÍ mirrors, which have not been recoated for two years and accumulated dust during dome work following telescope installation. Cleaning and recoating the mirrors are planned for September 2026 and are expected to improve the  sensitivity of TEQUILA.

\subsection{Calibration with Pupil Tracking}\label{ssec:pupil}

\begin{figure}
\centering
    \includegraphics[height=0.3\linewidth]{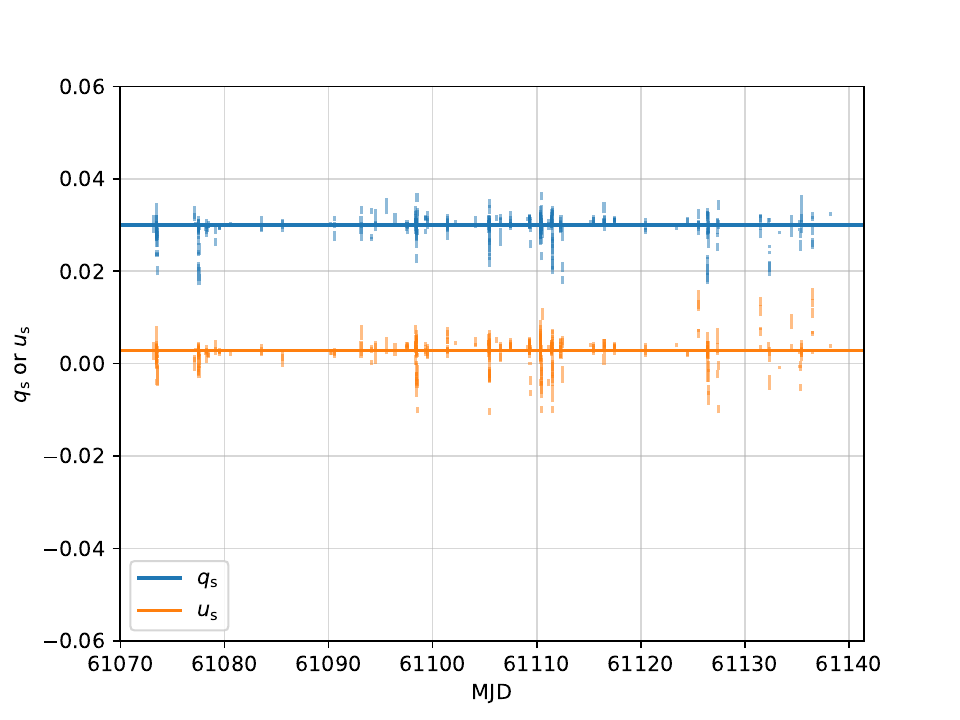}
    \includegraphics[height=0.3\linewidth]{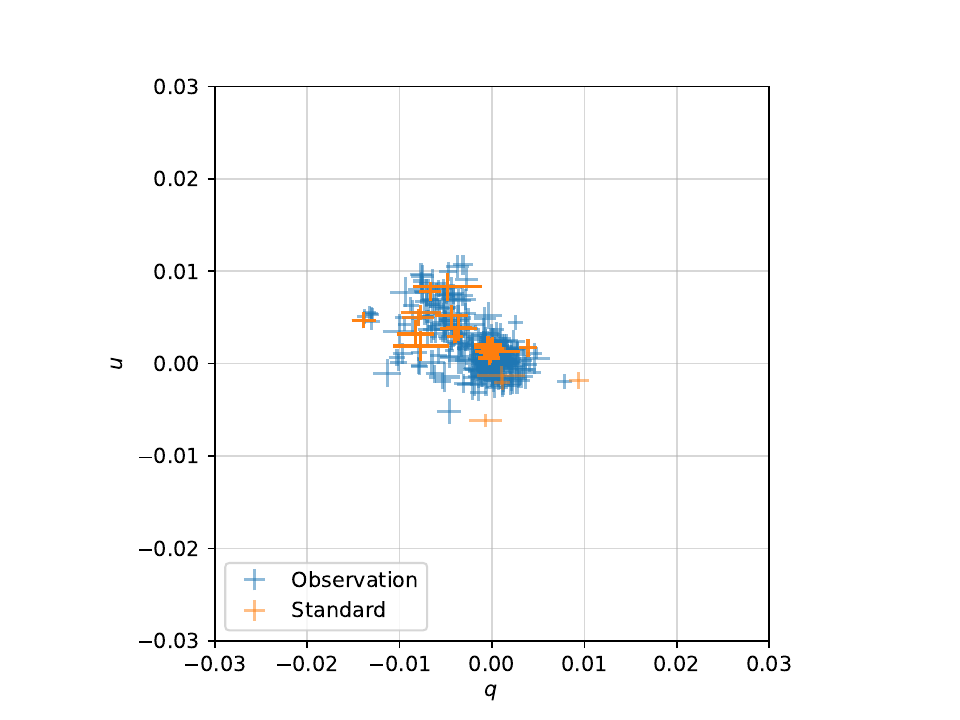}

    \includegraphics[height=0.3\linewidth]{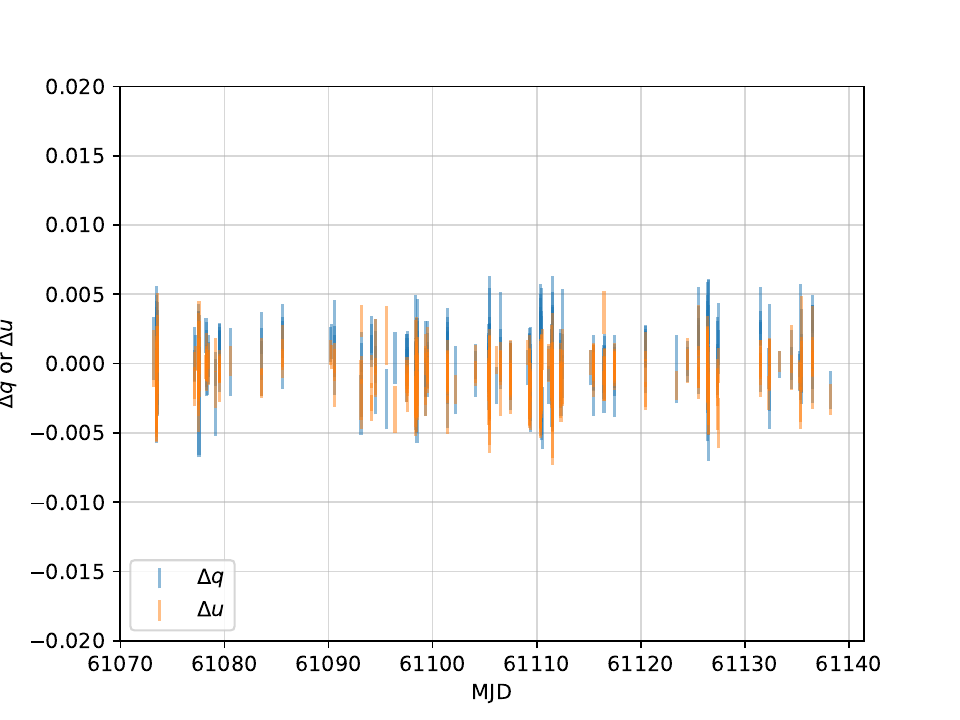}
    \includegraphics[height=0.3\linewidth]{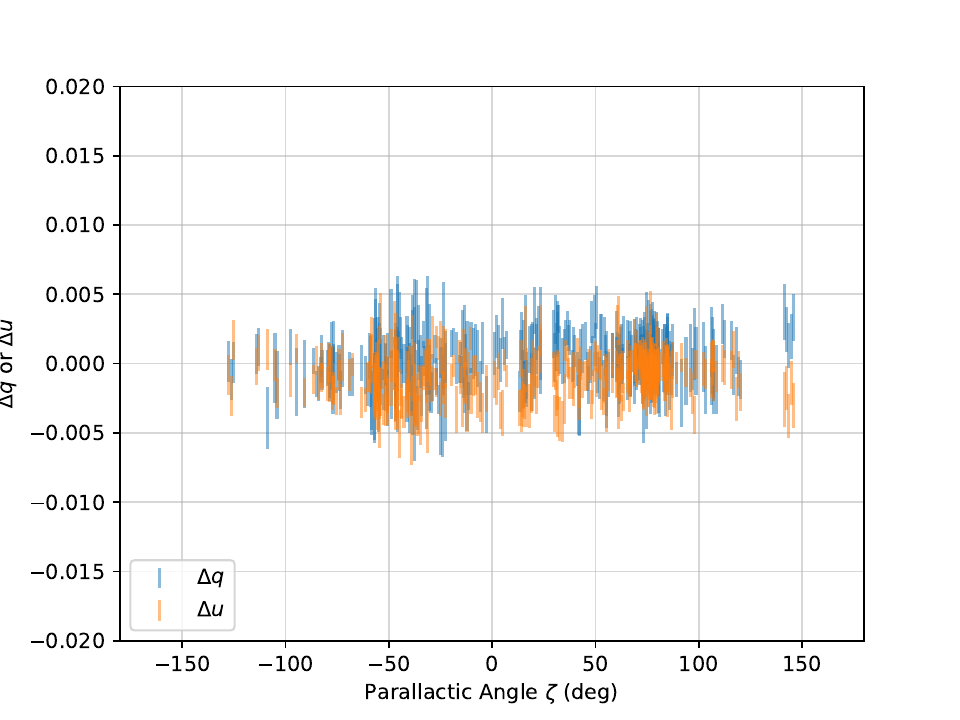}

    \caption{Calibration with pupil tracking. 
    Upper left: The raw polarizations $\qs$ and $\us$ for 403 observations of RoboPol standards. We see a strong instrumental polarization in $\qs$, a weaker one in $\us$, and dispersion due to the intrinsic polarization of the standards.
    Upper right: Values of $(q,u)$ for the standards and observations after transforming the observations from $(\qs,\us)$ to $(q,u)$.
    Lower left: Residuals $\Delta q \equiv q - q_0$ and $\Delta u \equiv u - u_0$ (observation minus standard) as a function of date.
    Lower right: Residuals $\Delta q$ and $\Delta u$ as a function of parallactic angle $\zeta$.
    }
    \label{fig:pupil-tracking-calibration}
\end{figure}

{\TEQUILA} is mounted on the Nasmyth focus of an altitude-azimuth telescope, so if the instrument is set to track the field the telescope mirror and the induced polarization from its tertiary mirror rotate with respect to the field and the instrument. 
We therefore began our investigation of the instrumental polarization using observations with pupil-tracking. Pupil tracking maintains a fixed orientation between the instrument and the tertiary mirror, which should lead to a more constant instrument polarization in the sensor frame, but the disadvantage is that the sensor is aligned on the sky with the parallactic angle plus a constant and rotates with respect to the field.

To implement pupil tracking, at the start of each exposure we rotate the derotator by an angle equal to the zenith distance plus a constant. We do not track the pupil through the exposure, but since our calibration exposures are short this is not a problem.

Our simple model to transform raw polarizations $(\qs,\us)$ to corrected polarizations $(q,u)$ for pupil-tracking observations has
\begin{align}
q &= (\qs - \Delta\qs) \cos (2(\zeta+\Delta\zeta)) - (\us - \Delta\us) \sin (2(\zeta+\Delta\zeta)),\mathand
u &= (\qs - \Delta\qs) \sin (2(\zeta+\Delta\zeta)) + (\us - \Delta\us) \cos (2(\zeta+\Delta\zeta)).
\end{align}
That is, we subtract a constant instrumental polarization $(\Delta\qs,\Delta\us)$ from the raw polarization and then transform to the celestial frame by rotating by an angle $\zeta+\Delta\zeta$, in which $\zeta$ is the parallactic angle and $\Delta\zeta$ allows for a misalignment between the sensor and the parallactic angle. 

In terms of Mueller matrices, this model could be represented by a rotation from the celestial frame to a parallactic frame, application of the combined matrix for the polarization of the telescope and the instrument (dominated by the reflection off the tertiary mirror), and then rotation back into the celestial frame.

We determine the values of $\Delta\qs$, $\Delta\us$, and $\Delta\zeta$ by minimizing the $\chi^2$ between the observed $(q,u)$ and standard $(q,u)$ for 403 observations of RoboPol standards \cite{2023A&A...677A.144B} on 41 nights between 2026 February 02 and 2026 April 08. The RMS uncertainty in these observations is 0.09\% in each of $\qs$ and $\us$ and the RMS uncertainty in the standard polarizations is 0.10\%. We find
\begin{align}
    \Delta\qs &= +2.98\%,\\
    \Delta\us &= +0.40\%,\mathand
    \Delta\zeta &=-8.38\deg.
\end{align}

The raw and calibrated data are shown in Figure \ref{fig:pupil-tracking-calibration}.
The RMS dispersion in $\Delta q$ and $\Delta u$ (observations minus standard) is 0.16\%. This dispersion is equal to that expected from the uncertainties in the observation and the standards. That is, to the limit of the uncertainties in the observations and the standards, there is no evidence for additional dispersion from imperfections in the model. This exercise demonstrates that {\TEQUILA} used with pupil-tracking is capable of producing polarimetry with RMS uncertainties of 0.15\% in fractional absolute polarization. Since uncertainties in the standards make a significant contribution to this, we suggest that the precision of {\TEQUILA} for relative polarimetry is likely to be at the level of 0.10\% or perhaps better.

\subsection{Calibration with Field Tracking}
\label{ssec:field}

\begin{figure}
\centering
    \includegraphics[height=0.3\linewidth]{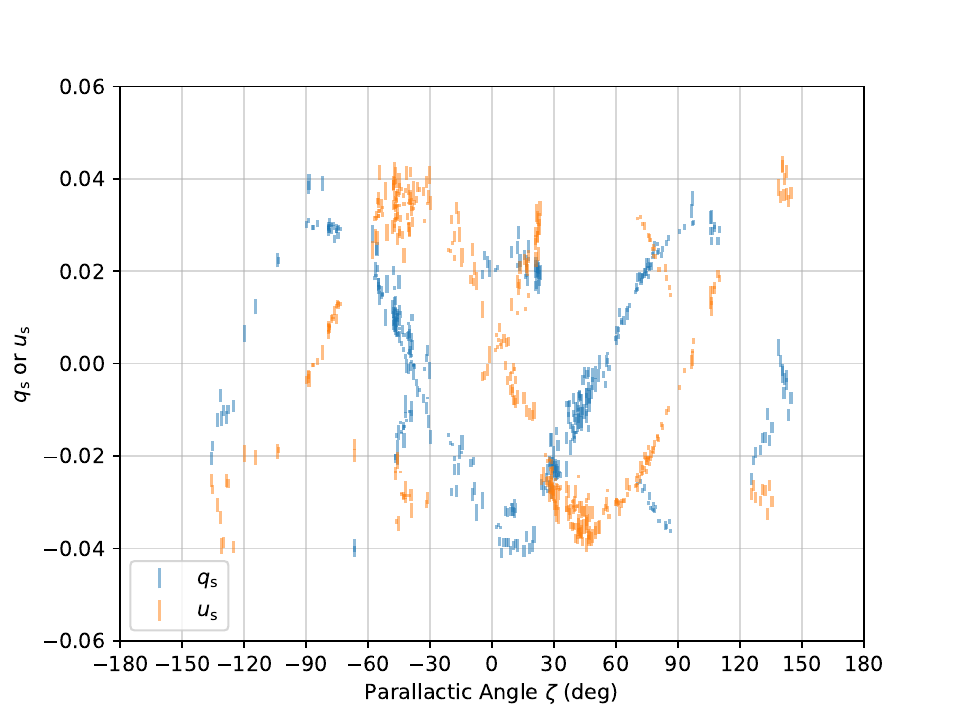}
    \includegraphics[height=0.3\linewidth]{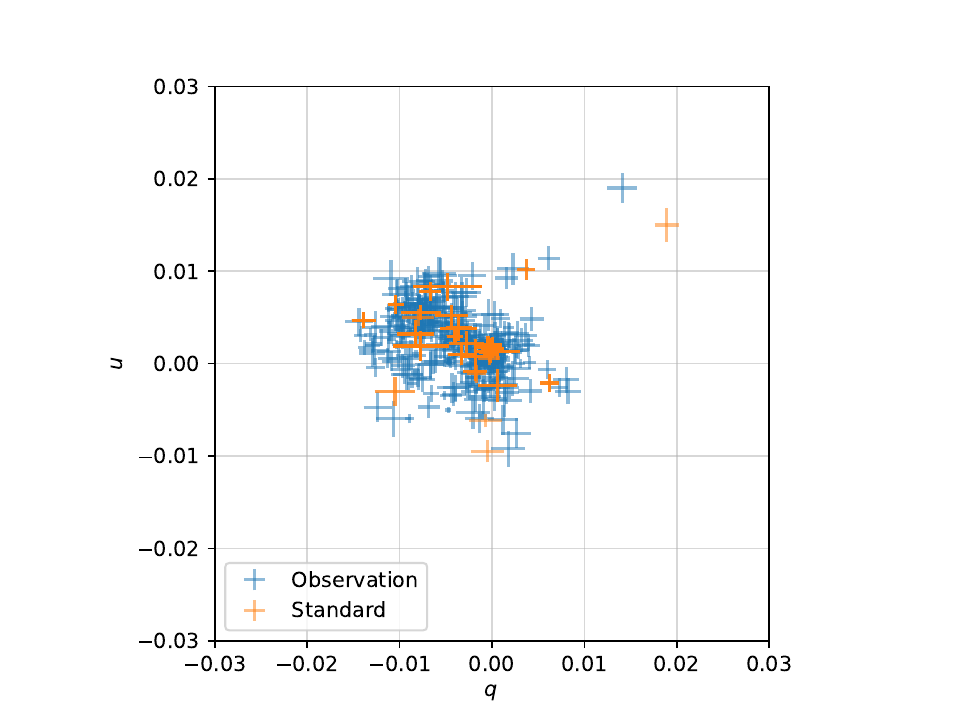}

    \includegraphics[height=0.3\linewidth]{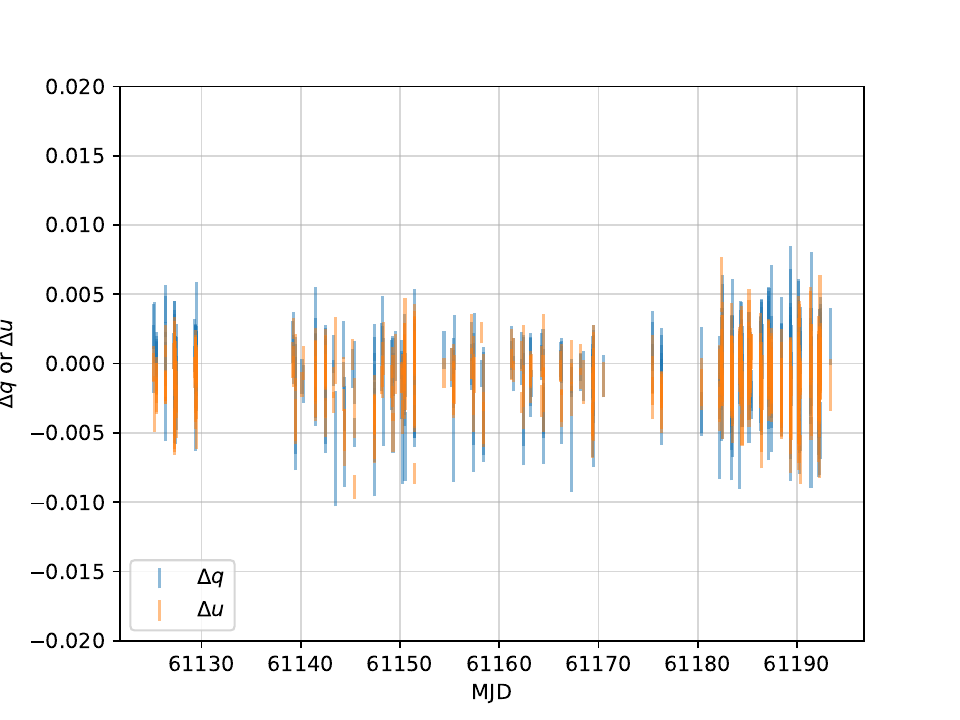}
    \includegraphics[height=0.3\linewidth]{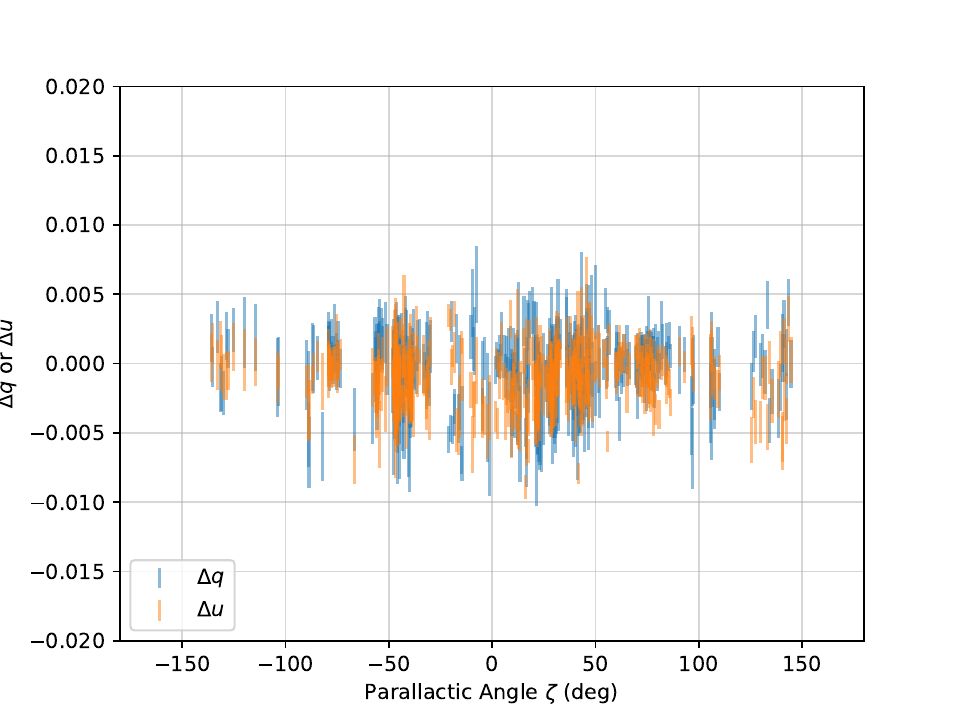}

    \caption{Calibration with field tracking. 
    Upper left: The raw polarizations $\qs$ and $\us$ for 579 observations of RoboPol standards. We see hints of the expected $\sin(2\zeta)$ and $\cos(2\zeta)$ modulation, but it is confused by the sign flips when tracking with offsets of $90\deg$ or $270\deg$, the corrections in the sensor frame which happen prior to these flips, and by the intrinsic polarizations of the standards.
    Upper right: Values of $(q,u)$ for the standards and observations after transforming the observations from $(\qs,\us)$ to $(q,u)$.
    Lower left: Residuals $\Delta q \equiv q - q_0$ and $\Delta u \equiv u - u_0$ (observation minus standard) as a function of date.
    Lower right: Residuals $\Delta q$ and $\Delta u$ as a function of the parallactic angle $\zeta$.
    }
    \label{fig:field-tracking-calibration}
\end{figure}

Observations with pupil tracking are fine for point sources, but we also want to use {\TEQUILA} for polarimetry of extended sources, and for these conventional field tracking is more convenient. Therefore, we also calibrated the instrument in this mode.
In this, we were aided by previous work on calibrating polarimeters at Nasmyth foci.\cite{2003SPIE.4843..456G,2007PASP..119.1371T,2018PASA...35...12W,2019ASSL..460...33B,2019MNRAS.482.5023H}

Field tracking for {\TEQUILA} is implemented unconventionally as the instrument does not currently have a cable wrap that allows it to safely turn through $360\deg$ on the derotator. Therefore, the sensor tracks the field during an exposure, but the rotation between the sensor and north can be an angle $\Delta\eta$ of $0\deg$, $90\deg$, $180\deg$, or $270\deg$ plus a constant offset $\Delta\theta'$ for imperfect alignment of the sensor on the sky. When the sensor is rotated by $90\deg$ or $270\deg$, the signs of $\qs$ and $\us$ are flipped compared to their values at $0\deg$ and $180\deg$.

Our simple model to transform raw polarizations $(\qs,\us)$ to corrected polarizations $(q,u)$ for field-tracking observations has
\begin{align}
\qs' &= (\wbox{$\us$}{$\qs$} - \Delta\wbox{$\us'$}{$\qs'$}) \cos(2\Delta\eta),\\
\us' &= (\us - \Delta\us') \cos(2\Delta\eta),\\
q &= \qs' \cos (2\Delta\theta') - \us' \sin (2\Delta\theta')-\pt'\cos(2(\zeta+\Delta\zeta)'),\mathand
u &= \qs' \sin (2\Delta\theta') + \us' \cos (2\Delta\theta')-\pt'\sin(2(\zeta+\Delta\zeta')).
\end{align}
That is, we subtract a constant sensor polarization $(\Delta\qs',\Delta\us')$ from the raw polarization, then rotate by an angle $\Delta\eta$ to a frame approximately aligned with the sky, then rotate by an angle $\Delta\theta'$ to the celestial frame, and finally subtract a telescope polarization $\pt'$ aligned with the parallactic angle $\zeta$ plus a constant offset $\Delta\zeta'$. 

In terms of Mueller matrices, this model could be represented by a rotation from the celestial frame to a parallactic frame, application of the matrix for the polarization of the telescope (dominated by the reflection off the tertiary mirror), rotation back into the celestial frame, and then application of the matrix for the instrument.

We determine the values of $\Delta\qs'$, $\Delta\us'$, $\Delta\theta'$, $\pt'$, and $\Delta\zeta'$ by minimizing the $\chi^2$ between the observed $(q,u)$ and standard $(q,u)$ for 434 observations of RoboPol standards \cite{2023A&A...677A.144B} on 44 nights between 2026 March 26 and 2026 June 02. The RMS uncertainty in these observations is 0.10\% in each of $\qs$ and $\us$ and the RMS uncertainty in the standard polarizations is 0.10\%. We find
\begin{align}
    \Delta\qs' &= -0.38\%,\\
    \Delta\us' &= +0.43\%,\\
    \Delta\theta' &=+6.93\deg,\\
    \pt' &= \phantom{+}3.45\%,\mathand
    \Delta\zeta' &=-0.05\deg.
\end{align}
We interpret the value of $\pt'$ as indicating that the tertiary mirror of the telescope induces about 3.5\% linear polarization and the small values of $\Delta\qs'$ and $\Delta\us'$, each less than 0.5\%, as indicative of errors in the flat field.
The small value of $\Delta\zeta'$ confirm that the telescope polarization is dominated by reflection off the tertiary mirror.
The raw and calibrated data are shown in Figure \ref{fig:field-tracking-calibration}.
The residuals have a standard deviation of about 0.22\%. This is slightly more than that expected from the combined uncertainties in the observation and the standards of 0.18\% and suggests that the transformation is contributing an RMS dispersion of about 0.10\%. 
Nevertheless, this exercise demonstrates that {\TEQUILA} used with field-tracking is capable of producing absolute polarimetry with RMS uncertainties of 0.20\%.

\begin{figure}
\centering
    \includegraphics[height=0.3\linewidth]{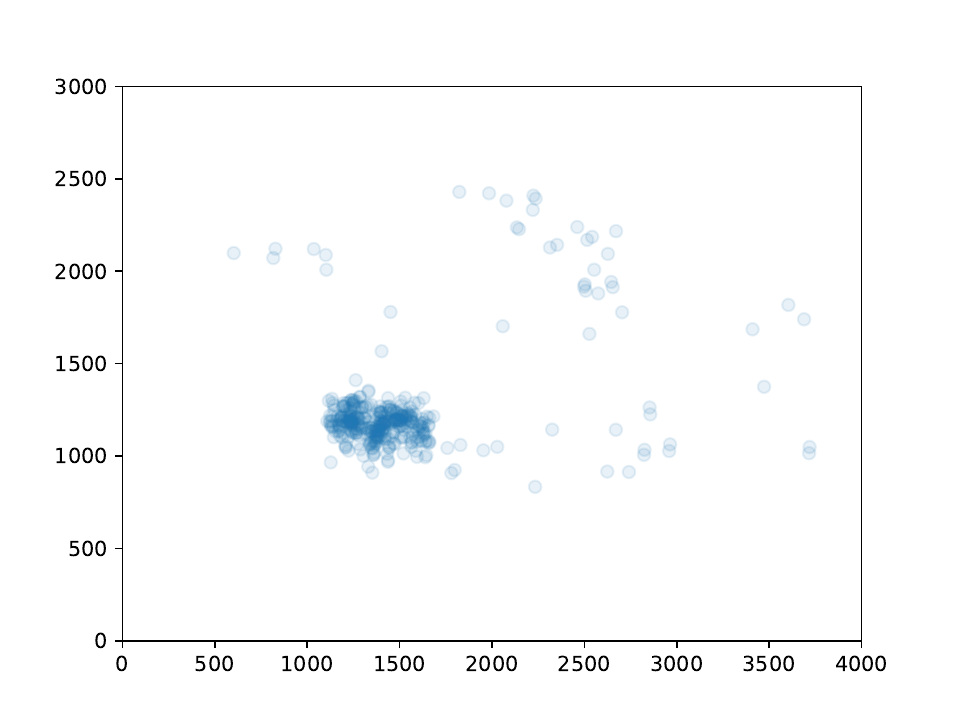}
    \includegraphics[height=0.3\linewidth]{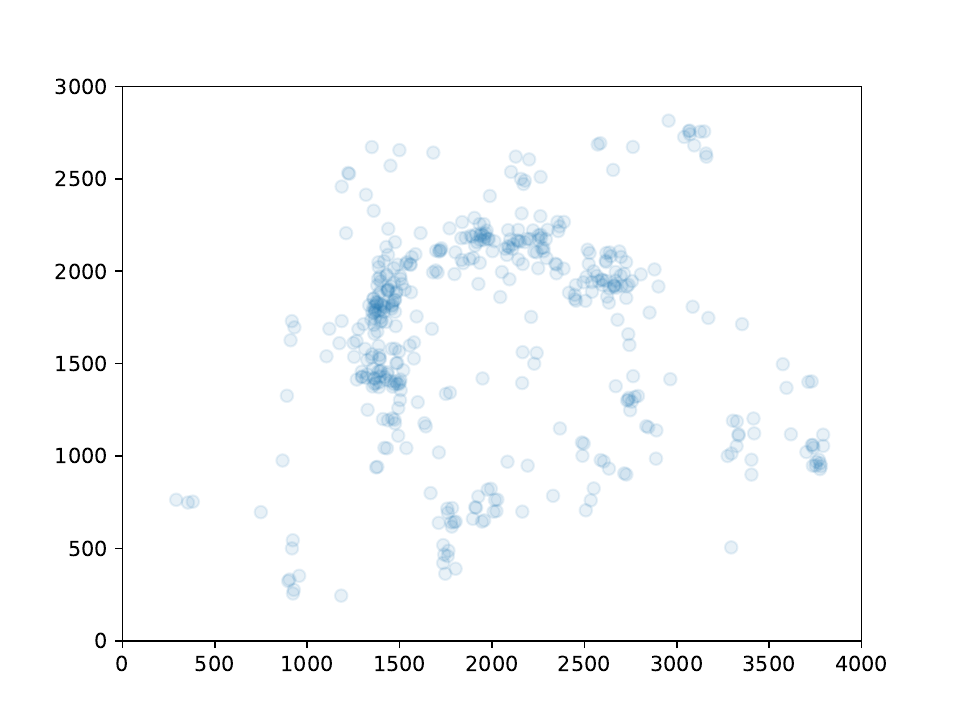}
    \caption{The centers of the observations of each group of standards for the pupil-tracking calibration (left) and field-tracking calibration (right). The wider dispersion of the standards in the focal plane make the field-tracking calibration more susceptible to flat-fielding errors.
    }
    \label{fig:calibration-xy}
\end{figure}

The slightly worse performance in field-tracking observations is perhaps expected for three reasons. First, in the field-tracking observations the telescope polarization rotates with respect to the instrument and field, giving a less stable configuration. Second, as we show in section \ref{sec:flat-fields} below, the instrumental polarization in the flat-field depends on the derotator angle, and we currently do not account for this in our reduction. Third, as Figure \ref{fig:calibration-xy} shows, as the pointing accuracy is worse in field-tracking mode than in pupil-tracking mode and so the greater dispersion in the focal plane makes the field-tracking calibration more susceptible to flat-fielding errors.

\subsection{Consistency of the Two Calibration Models}

For observations on the meridian,  the instrument, telescope, and field geometries are the same for both pupil-tracking and field-tracking observations, and so we expect the transformations from $(\qs,\us)$ to $(q,u)$ to be equivalent.
The pupil-tracking transformation has
\begin{align}
\wbox{$\us$}{$\qs$} - \wbox{$u$}{$q$}  &\approx \Delta\wbox{$\us$}{$\qs$} \approx 2.98\%,\mathand
\us - u  &\approx \Delta\us \approx 0.40\%,
\end{align}
in which we have ignored the small misalignment terms in  $\sin(2\Delta\zeta)$.
The field-tracking transformation has
\begin{align}
\wbox{$\us$}{$\qs$} - \wbox{$u$}{$q$}   &\approx \Delta\qs' + \pt' \approx -0.38\% +3.45\% = +3.07\%,\mathand
\us - u &\approx \wbox{$\Delta\qs' + \pt'$}{$\Delta\us'$} \approx +0.43\%,
\end{align}
in which we have ignored the small misalignment terms in $\sin(2\Delta\theta')$ and  $\sin(2\Delta\zeta')$.
We see the predictions of both models for $\qs - q$ ($2.98\%$ and $3.07\%$) and $\us-u$ ($0.40\%$ and $0.43\%$) agree at the level of 0.1\%.

\section{Flat-Field Polarization Structure}
\label{sec:flat-fields}

In this section, we analyze the polarization structure of the flat-field images.
We obtained flat-field images in the laboratory and also twilight flat-field images on the telescope. 
As mentioned previously, we normalize each channel separately from the median values in the central 50\% of the rows and columns to compensate for linear polarization in the illuminating sources.
Here, we analyze the polarization images in $q$ and $u$ formed by applying equations \ref{eq:ideal-q} and \ref{eq:ideal-u} to each super-pixel in the flat-field images.

The flat-field polarization images exhibit smooth, large-scale spatial variations that we model using a two-dimensional expansion in Zernike polynomials. The sensor coordinates are first centered on the sensor midpoint and normalized by half of the sensor dimensions before being transformed into polar coordinates,
$r=\sqrt{x^2+y^2}$, 
$\theta=\operatorname{atan2}(y,x)$, 
which are used to construct a complete Zernike basis up to radial order $n_{\rm max}$. That is, the unit disk over which Zernike polynomials are normally used maps to the ellipse inscribed in the sensor.

The usual Zernike polynomials\cite{1976JOSA...66..207N} $Z_n^m$ are defined by
\begin{equation}
R_n^m(r)=
\sum_{k=0}^{(n-|m|)/2}
(-1)^k
\frac{(n-k)!}
{k!\left(\frac{n+|m|}{2}-k\right)!
\left(\frac{n-|m|}{2}-k\right)!}
r^{\,n-2k},
\end{equation}
and
\begin{equation}
Z_n^m(r,\theta)=
\begin{cases}
R_n^m(r)\cos(m\theta), & m>0,\\
R_n^{|m|}(r)\sin(|m|\theta), & m<0,\\
R_n^0(r), & m=0.
\end{cases}
\end{equation}

Although the usual Zernike polynomials form an orthogonal basis on the unit disk, in this work the basis functions are evaluated over the entire sensor without masking pixels outside the unit circle. Consequently, the analytical orthogonality of the Zernike polynomials is not preserved on the sampled sensor grid. Each basis function is numerically normalized to have unit root-mean-square (RMS) amplitude over the sampled sensor pixels,
which gives us modified Zernike polynomials $\hat{Z}_n^m$ defined by
\begin{equation}
\hat{Z}_n^m(x,y)=
\frac{Z_n^m(x,y)}
{\left[\displaystyle\frac{1}{N}\sum_{i=1}^{N}
\left(Z_n^m(x_i,y_i)\right)^2\right]^{1/2}},
\end{equation}
so that 
\begin{equation}
\frac{1}{N}\sum_{i=1}^{N}
\left(
\hat{Z}_n^m(x_i,y_i)
\right)^2
=1.
\end{equation}
This normalization allows the fitted coefficients to be interpreted directly as the RMS amplitudes of the corresponding spatial modes over the sampled sensor.

The observed polarization maps are modeled through a linear least-squares fit,
\begin{equation}
\mathbf{c}
=
\arg\min_{\mathbf{c}}
\left\|
\mathbf{A}\mathbf{c}-\mathbf{y}
\right\|^2,
\end{equation}
where $\mathbf{y}$ is the vector of measured polarization values ($q$ or $u$), $\mathbf{c}$ is the vector of fitted Zernike coefficients, and $\mathbf{A}$ is the design matrix whose columns contain the Zernike basis functions evaluated at the sensor coordinates,
\begin{equation}
\mathbf{A}=
\begin{bmatrix}
\hat Z_1(r_1,\theta_1) & \hat Z_2(r_1,\theta_1) & \cdots & \hat Z_M(r_1,\theta_1)\\
\hat Z_1(r_2,\theta_2) & \hat Z_2(r_2,\theta_2) & \cdots & \hat Z_M(r_2,\theta_2)\\
\vdots & \vdots & \ddots & \vdots\\
\hat Z_1(r_N,\theta_N) & \hat Z_2(r_N,\theta_N) & \cdots & \hat Z_M(r_N,\theta_N)
\end{bmatrix},
\end{equation}
where $N$ is the number of sensor pixels and $M$ is the number of Zernike basis functions. 
The model image is then given by
\begin{align}
q(x,y)
&=
\sum_{n,m}
c_{nm}^{(q)}
\,
\hat{Z}_n^m(x,y),
\mathand
u(x,y)
&=
\sum_{n,m}
c_{nm}^{(u)}
\,
\hat{Z}_n^m(x,y).
\end{align}
The least-squares solution determines the coefficients that minimize the sum of squared residuals between the measured polarization images and their Zernike reconstructions. The reconstructed instrumental polarization images are obtained from
$\mathbf{y}_{\rm model} = \mathbf{A}\mathbf{c}$,
and residuals are computed as
$\mathbf{y}_{\rm resid} = \mathbf{y} -
\mathbf{y}_{\rm model}$.
From the model $q$ and $u$ images we compute the degree $p$ and angle $\theta$ of linear polarization using equations \ref{eq:p} and \ref{eq:theta}, and use these  to visualize the instrumental polarization across the sensor.

\subsection{Flat-Field in the Laboratory}

\begin{figure} [tp]
\centering
   \includegraphics[height=10.5cm]{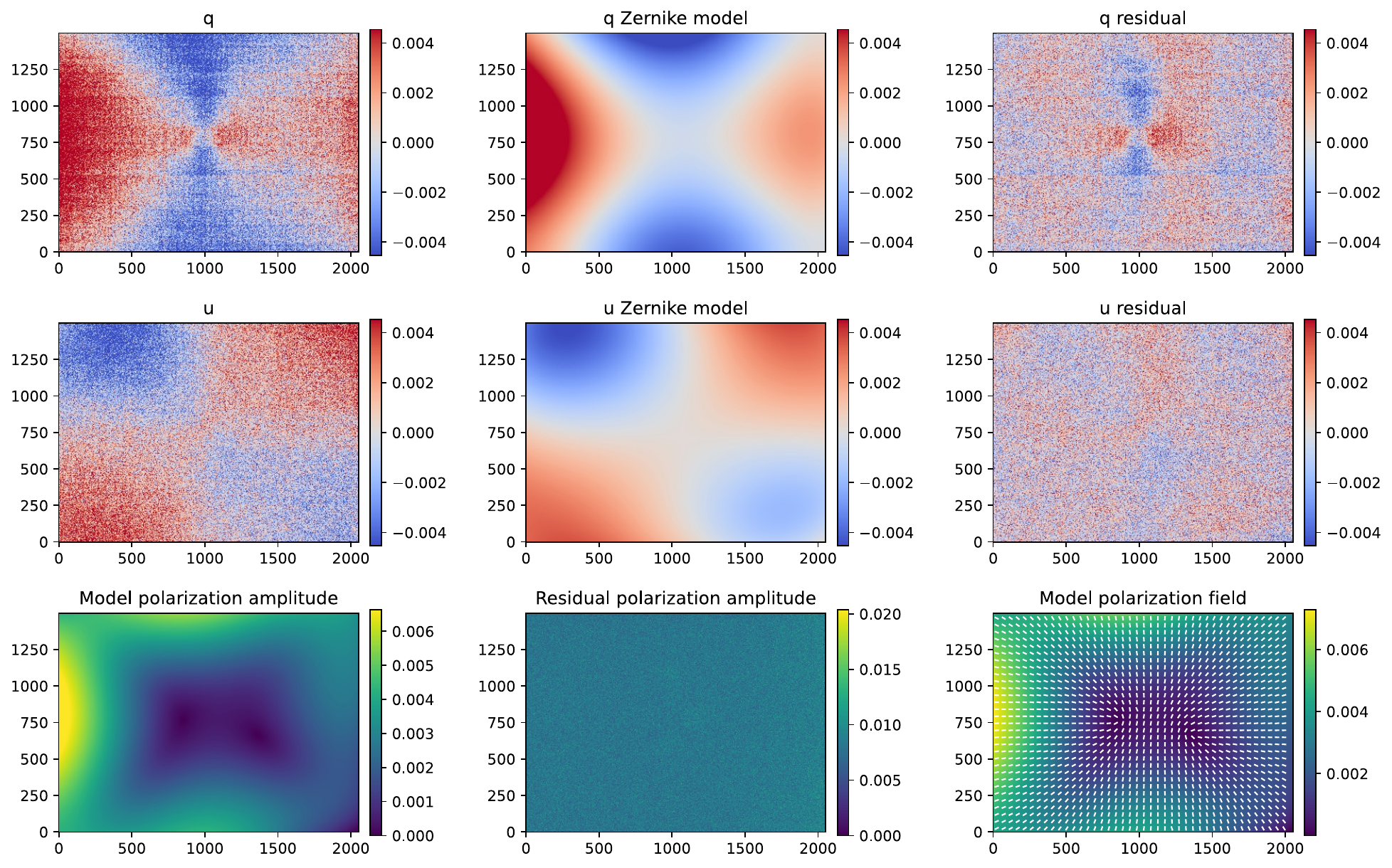}
   \caption{Instrumental polarization derived from the laboratory flat-field image obtained with collimated light using a Zernike decomposition with $M=15$ polynomials. The panels show the measured $q$ and $u$ polarization images, the corresponding Zernike reconstructions, the residuals, and the resulting degree and orientation of linear polarization. The laboratory flat exhibits the same characteristic radial pattern observed in the on-sky flat fields, suggesting that the effect is intrinsic to the instrument rather than the illumination source. The physical origin of this pattern is currently unknown, but it may be associated with small-scale non-planarity of the micro-polarizer array or with polarization introduced by window or filter.}   
   \label{fig:labflat}
\end{figure}

Figure \ref{fig:labflat} shows the polarization structure in a flat field taken in the laboratory with the instrument illuminated with an approximately collimated beam.
Here we have only the sensor, the window, and the filter. 
Increasing the maximum radial order of the expansion produced negligible improvement in the residuals, with the RMS residual remaining approximately $7\times10^{-3}$ for both $q$ and $u$. This indicates that the dominant instrumental polarization is well represented by low-order spatial modes, while the remaining signal consists primarily of higher-spatial-frequency structure, most likely pixel-to-pixel gain variations. 
The dominant low-order structure is a radial polarization that increases from zero in the center to about 0.005 (i.e., 0.5\%) at the edge of the sensor, with some regions being slightly higher. The normalization of the flats forces the polarization to be zero on average in the central region.

\subsection{Flat-field on the Telescope}

We repeated this analysis on twilight sky flat-field images to assess the stability of the instrumental polarization under on-sky observing conditions. These flats are taken with an $f/7.2$ converging beam. The twilight sky is highly polarized and the reflection from the tertiary also contributes, but as we have noted we normalize each channel separately from the median values in the central 50\% of the rows and columns which compensates for this and forces the average polarization in the central region to be zero.

\begin{figure} [tp]
\centering
      \includegraphics[height=10.5cm]{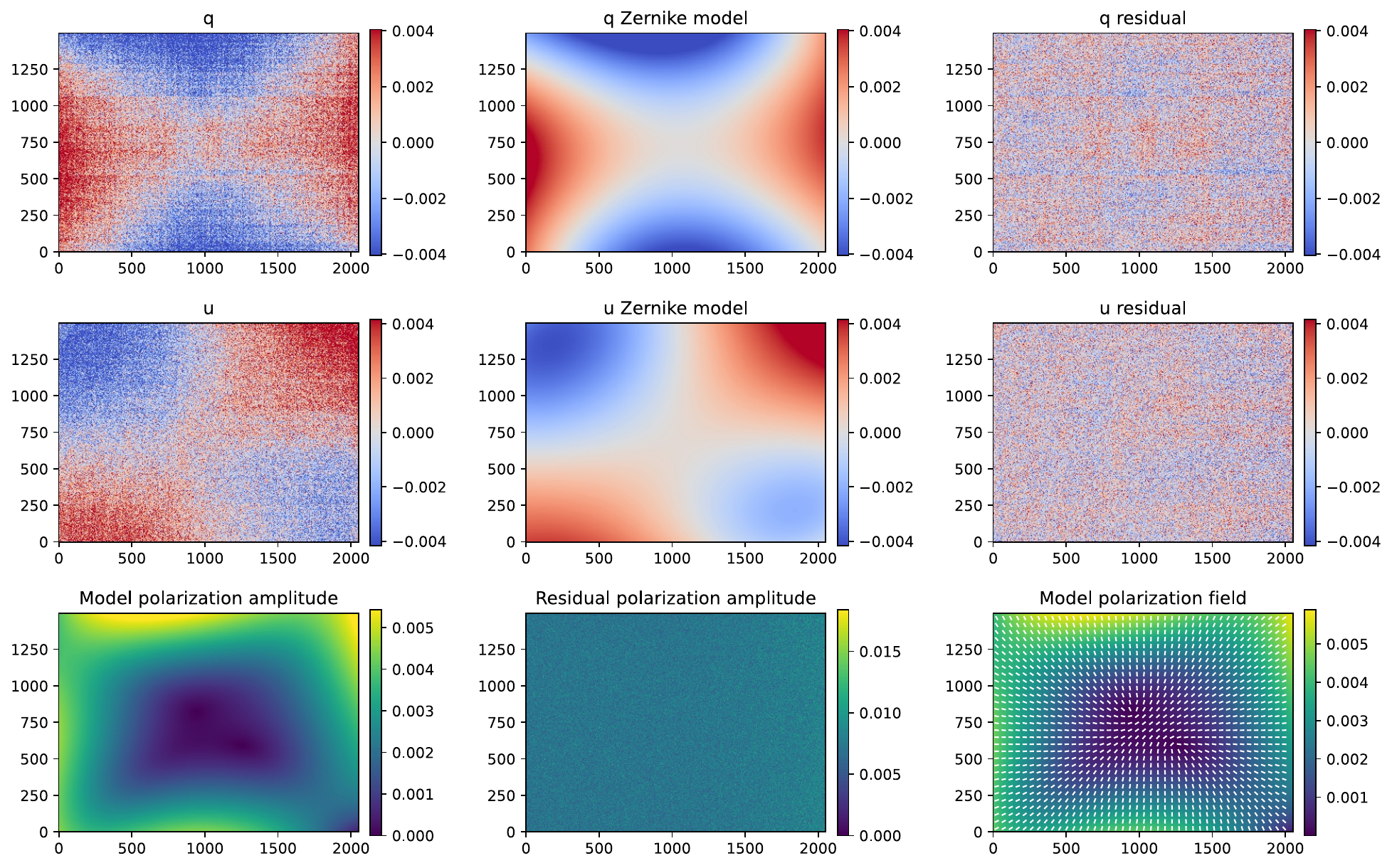}
   \caption{Instrumental polarization derived from the twilight-sky flat-field image obtained on the telescope and with the derotator at $-45\deg$ using a Zernike decomposition with $M=15$ polynomials. The panels are the same as in Figure \ref{fig:labflat}. The twilight-sky flat field exhibits the same characteristic radial pattern observed in the laboratory flat field, suggesting that the effect is intrinsic to the instrument rather than the illumination source.}   
   \label{fig:skyflat}
\end{figure}

Figure \ref{fig:skyflat} shows the analysis for a flat taken with the derotator at $-45\deg$.
The resulting polarization images exhibit behavior remarkably similar to those obtained from laboratory flat field, both in direction and amplitude. While the individual Zernike coefficients differ, perhaps reflecting differences in the illumination geometry, the overall amplitudes of the fitted and residual components are nearly unchanged.

\begin{figure} [tp]
\centering
    \includegraphics[height=5.5cm]{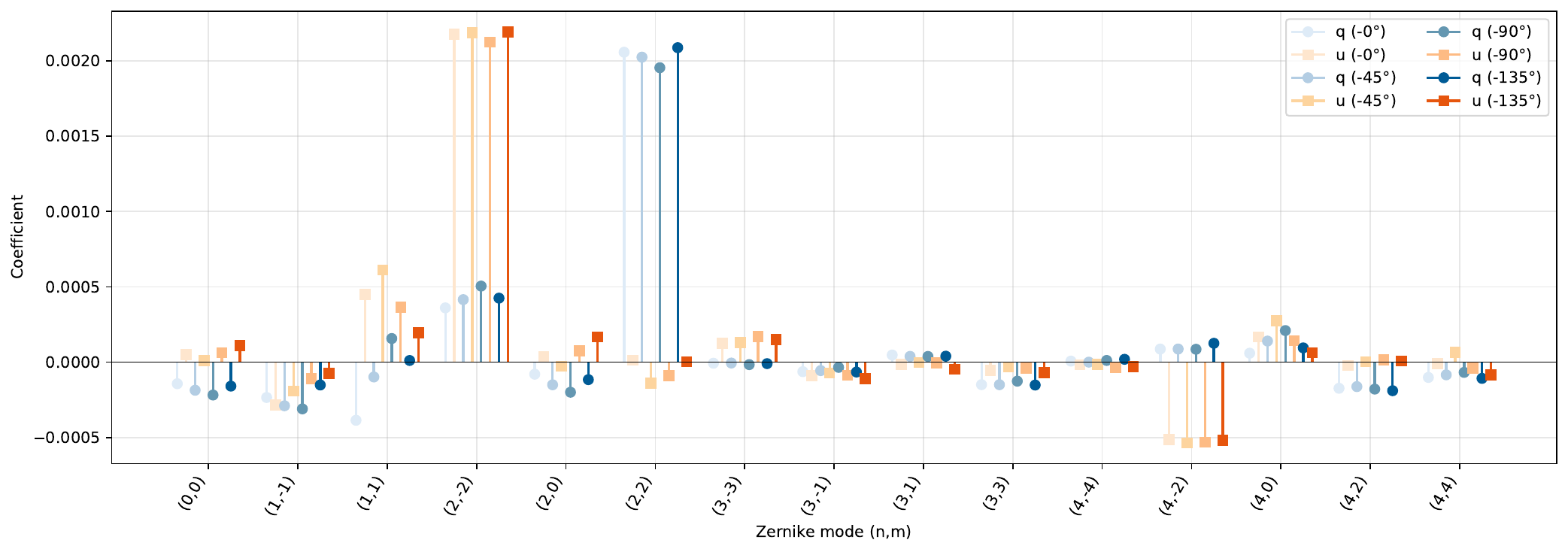}
   \caption{Comparison of the low-order Zernike coefficients describing the instrumental polarization measured  from flat-field observations obtained on sky, at derotator orientations of $0\deg$, $-45\deg$, $-90\deg$, and $-135\deg$.  The dominant contributions arise from the second-order modes, particularly ($n,m$)=($2,\pm2$), while higher-order coefficients remain comparatively small. The large-scale instrumental polarization pattern is largely independent of derotator orientation.}   
   \label{fig:skyflatzernike}

\bigskip
   \includegraphics[height=10.5cm]{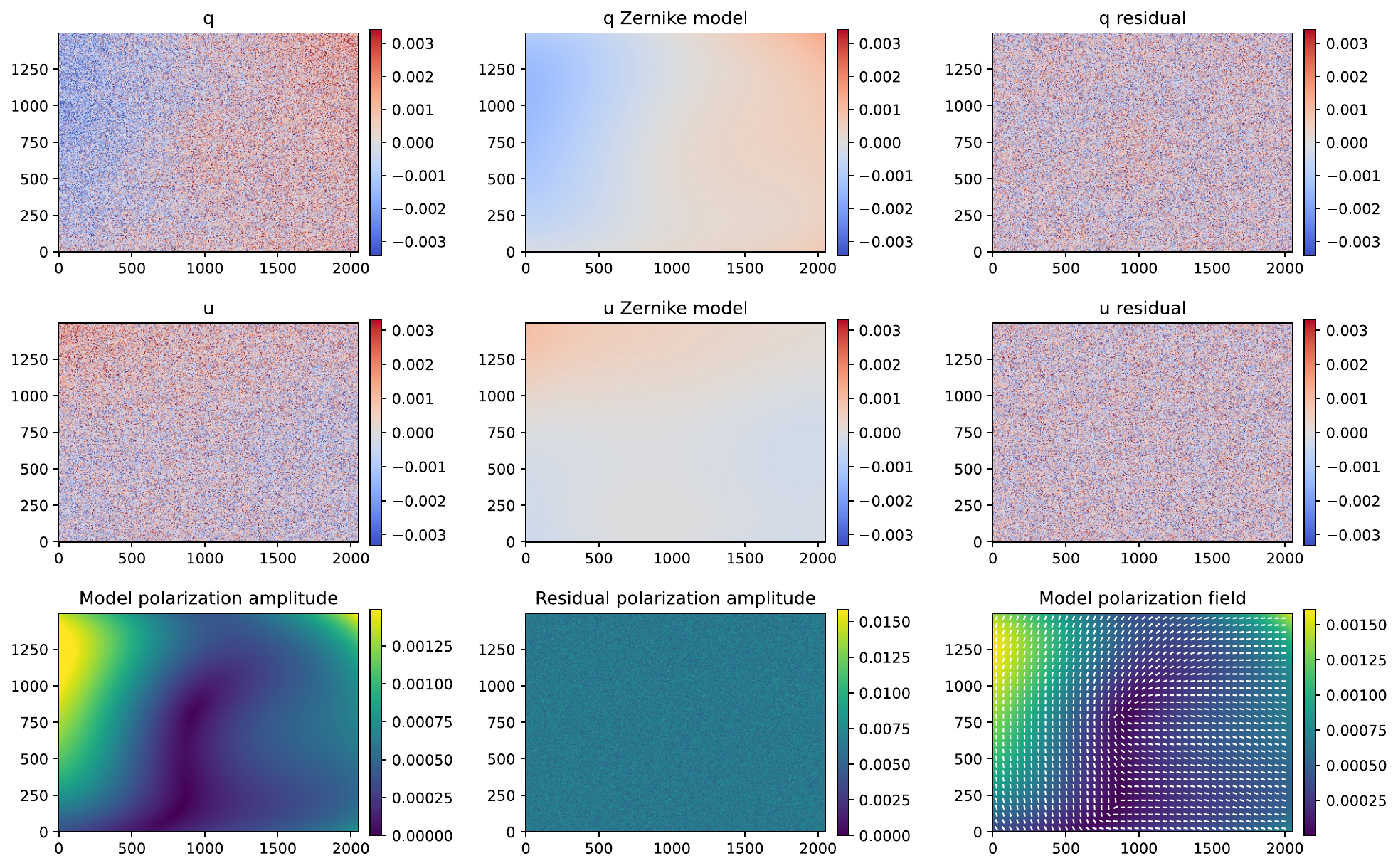}
   \caption{Differential instrumental polarization derived from the ratio of twilight flats on the telescope at $0\deg$ and $-90\deg$. The radial polarization is not dominant in the ratio, indicating that it does not change with the instrument rotation. Instead, the dominant terms correspond to gradients in $q$ and $u$ and are at the level of 0.0005 (0.05\%) RMS or $\pm0.001$ ($\pm0.1\%$) peak-to-valley.}   
   \label{fig:flatdivision}
\end{figure}

During field-tracking observations, the derotator rotates between $0\deg$ and slightly more than $-90\deg$. We therefore checked the stability of the twilight flat fields at $0\deg$, $-45\deg$, $-90\deg$, and $-135\deg$.
The Zernike decompositions obtained at these derotator angles are shown in Figure \ref{fig:skyflatzernike} and exhibit a remarkably stable instrumental polarization pattern dominated by a radial pattern. In all four configurations, the $q$ field is dominated by the $Z_{2,+2}$ mode, while the $u$ field is dominated by the $Z_{2,-2}$ mode. The amplitudes of these dominant coefficients vary by only approximately $0.0005$, demonstrating that the principal radial polarization structure is nearly invariant with derotator orientation and that the flat fields are expected to be stable under rotation at the level of 0.0005 (0.05\%) RMS.

We have also performed a polarization analysis on the ratio of the $0\deg$ and $-90\deg$ flats. The results are shown in Figure \ref{fig:flatdivision}. The structure in the ratio is at the level of 0.0005 (0.05\%) RMS or $\pm0.001$ ($\pm0.1$\%) peak-to-valley.

\subsection{Origin of the Polarization Structure}

The radial polarization structure seen in the twilight flat-fields could be due to the telescope, scattered light in the instrument, the filter, the window, the sensor, or some combination of some or all of these.

We would expect light reflected off the inner surfaces of the cylindrical tubes that attach the sensor package to the OGSE flange would lead to an axisymmetric pattern, but with tangential polarization. The reflection from the tertiary mirror might give slightly different polarizations for different field angles, but we would expect this to be a simple gradient over the sensor field --- and we might be seeing hints of this in the ratio of the flat fields shown in Figure \ref{fig:flatdivision} --- but again would not explain the radial pattern.

When light passes through an inclined plate, the reflected light becomes partially $s$ polarized, and the transmitted light  becomes partially $p$ polarized. In a non-telecentric optical system like COLIBRÍ, this can lead to a radial pattern of instrumental polarization. In TEQUILA on the telescope, this could be occurring in the sensor package window and the filter. However, we note that a very similar pattern is seen in the laboratory flat field, which was taken with a collimated beam and so should not show this effect. In order to dispel doubts, we plan to model the polarization introduced by the filter and window to determine if they can produce the observed amplitude and will repeat the laboratory flats with different degrees of collimation.

The last possibility is that this polarization pattern is intrinsic to the sensor. This would naturally explain why such similar patterns are seen in the laboratory and on the telescope. We have also examined the flat fields of a second sensor kindly provided to us by Yannis Liodakis, and they show similar structure. That said, it is not clear to us what feature of the sensor micro-lens array or micro-polarizer array might produce this effect.

\section{Remaining Calibration Issues}
\label{sec:remaining-calibration-issues}

One remaining issue is that we have not established uncertainties and correlations between our fitted parameters.
We will address this in a future iteration by fitting the parameters using Markov-chain Monte-Carlo methods. We suspect the worst uncertainty in both of our calibration models will be in the position angle of the polarization ($\Delta\eta$ or $\Delta\theta'$); more observations of high-polarization standards will improve this.

A second remaining issue is the additional calibration uncertainty in field-tracking observations. Improving the telescope pointing model so that we can use the same part of the sensor might well reduce this calibration uncertainty for point sources.
We will also consider taking into account the variation in the flat field with derotator position.
A third is 
that we have quantified the polarimetric uncertainty for point sources, but not yet for resolved sources. 
Understanding flat-field structure discussed in the previous section is relevant for having confidence in the polarimetry of point-sources at different positions in the field and of resolved sources.

\section{First Science Results}
\label{sec:first-science}

\begin{figure} [!h]
\centering
   \centering
   \raisebox{0.15cm}{%
      \includegraphics[height=6.6cm]{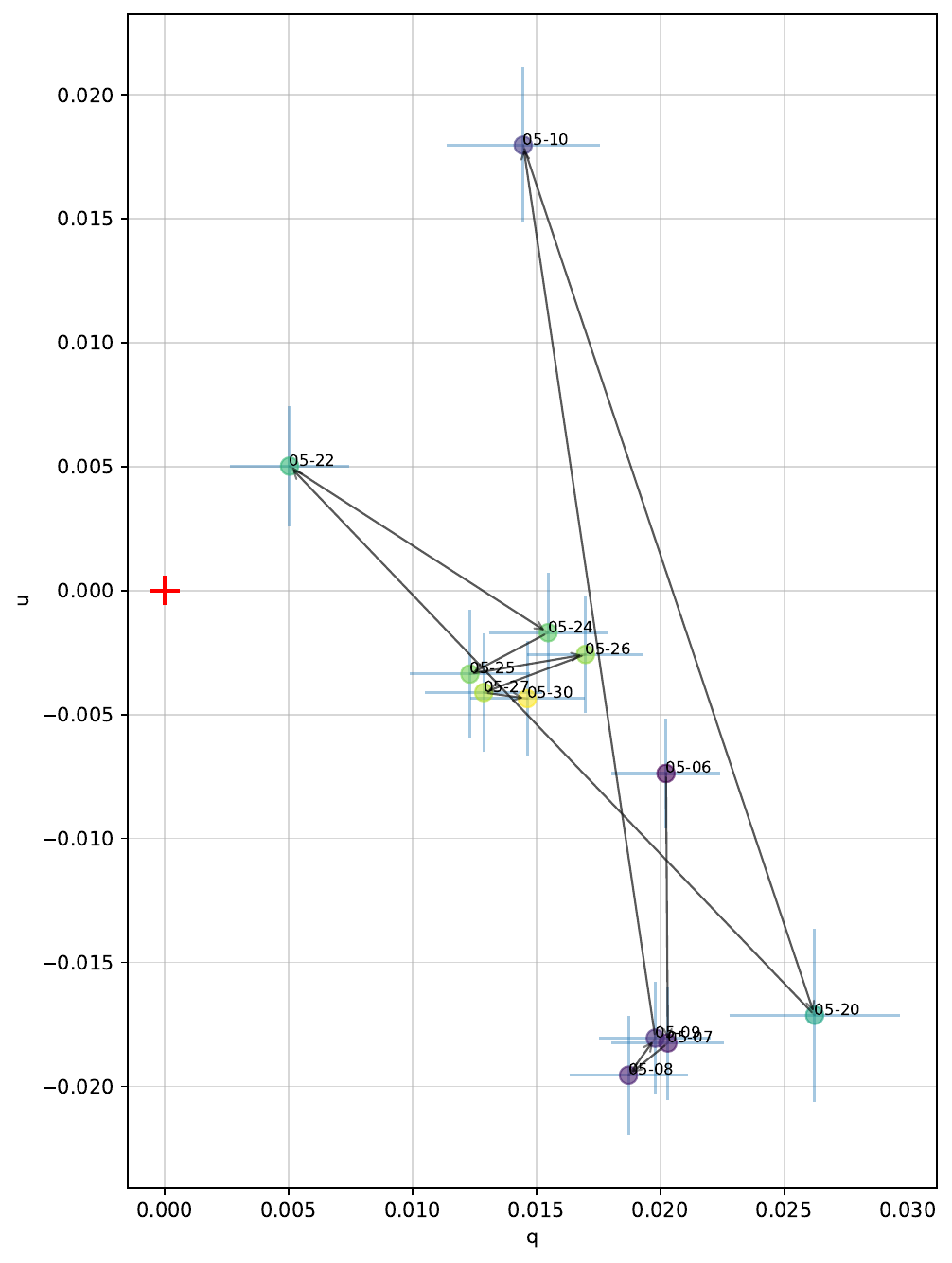}%
   }
   \includegraphics[height=7.0cm]{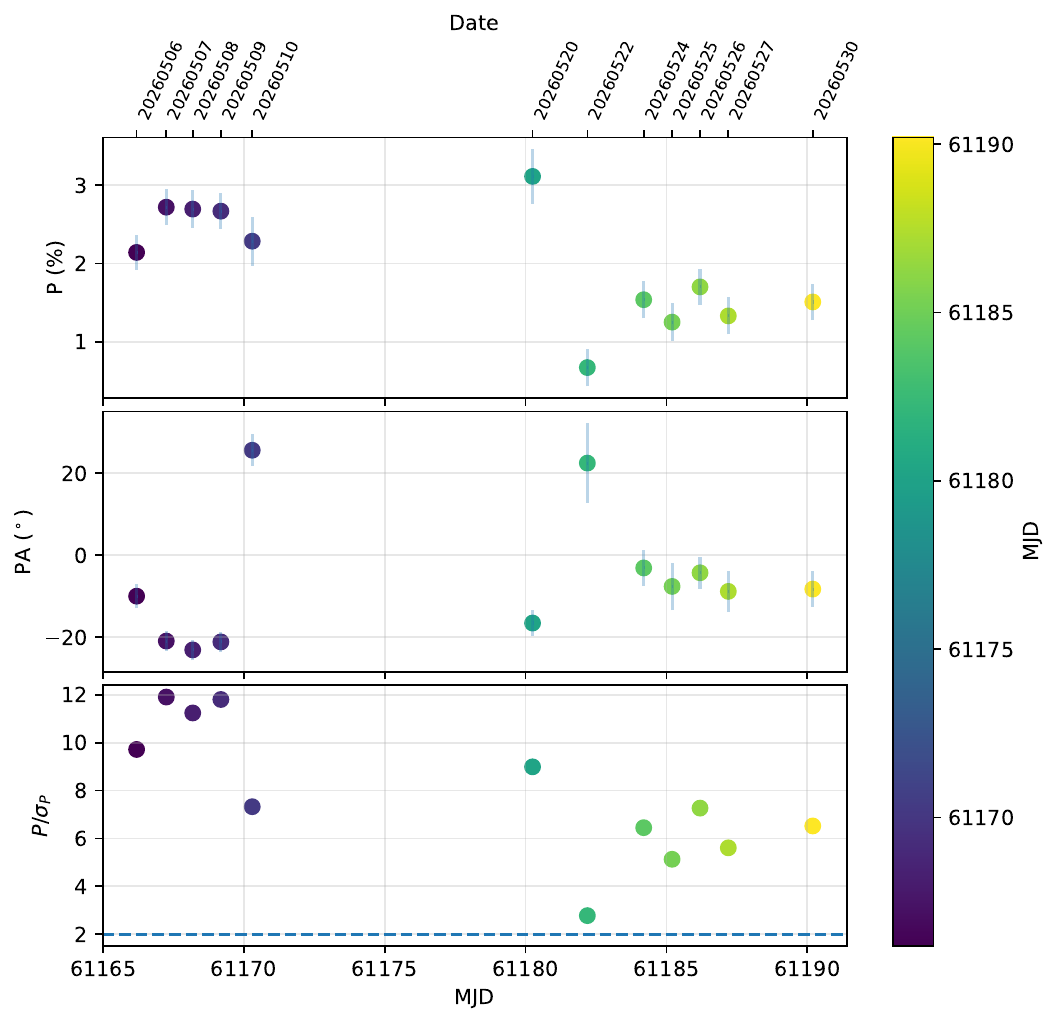}
   
   \caption{Left panel: Stokes parameters $q$ and $u$ are plotted with observational uncertainties and color-coded by Modified Julian Date (MJD) to represent temporal evolution. Successive observations are connected to illustrate the polarization vector motion through the $q$–$u$ plane. The red cross marks the origin. Right top panel: fractional polarization degree $p$ (\%). Right middle panel: polarization angle (PA, degrees). Right bottom panel: polarization signal-to-noise ratio. Error bars represent propagated uncertainties. A dashed horizontal line in the bottom panel marks the SNR threshold of 2. The top axis provides corresponding calendar dates for reference.}   
   \label{fig:Mkr421}
\end{figure}

Markarian 421 is one of the nearest and brightest blazars, making it a key target for X-ray polarimetry \cite{Liodakis2019} and a laboratory for studying relativistic jets and high-energy particle acceleration. Previous joint X-ray and optical polarimetric observations, including measurements from IXPE, have provided evidence consistent with an energy-stratified jet in which polarization varies with photon energy \cite{Bharathan2024}. Further progress requires coordinated multi-wavelength polarimetric monitoring, and {\TEQUILA} has allowed us to participate in such an effort by providing regular optical measurements of the polarization degree and angle over the past several months. The observations shown here represent only a small fraction of the data collected and primarily serve to demonstrate the scientific utility of our instrument.

Our results are shown in Figure~\ref{fig:Mkr421}. We detect a persistent optical polarization of up to approximately 3\%, along with significant variability in the polarization position angle. The polarization behavior shows distinct clustering in $q$–$u$ space, with repeated transitions between preferred states.

\section{Summary and Discussion}
\label{sec:pontification}

We have presented the design, calibration, and early science results for the {\TEQUILA} imaging polarimeter. We have shown that we can determine the linear polarization of point sources with uncertainties of about 0.15\%--0.20\%, although we are still working to understand and calibrate uncertainties in the flat fields. Applications of these sensors have already been explored in fields such as atmospheric remote sensing \cite{Stockmans2024}. Here, we have shown that they can also enable sub-percent astronomical polarimetry.

We note that we have demonstrated these results at the Nasmyth focus of an altitude-azimuth telescope, which is a challenging location for polarimetry. That we were able to obtain polarimetry apparently limited by photon noise and standard calibrations to the level of 0.15\% using pupil tracking is a positive sign for the precision of polarimetry with these devices at the conventional Cassegrain focus of equatorial telescopes.

While we have focused here on point sources, {\TEQUILA} also allow polarimetry of extended targets such as nebulae, and we will present imaging polarimetry of such objects in a forthcoming paper.

Our experience directly demonstrates that wire-grid micro-polarizer sensors open astronomical polarimetry in ways that were previously unthinkable. 
We had not previously built a polarimeter, but we were able to create an imaging polarimeter from commercial components costing about US\$15,000 and start science observations in about one year of part-time effort.
Comparable astronomical polarimeters have traditionally required years of development, dedicated engineering support, and budgets at least an order of magnitude larger. This possibility represents a transformative new opportunity for astronomical polarimetry, placing imaging polarimeters within reach of many university observatories and, increasingly, experienced citizen astronomers.

The principal remaining challenge is calibration. Developing widely applicable software tools to automate much of the reduction and calibration process---a polarimetric analog of STDPipe \cite{Karpov2021}---would substantially lower the barrier to entry for astronomical polarimetry.

An intriguing possibility, raised in conversations with Yair Krongold, Julio Ramírez-Vélez, David Hiriart, and Roger Blandford, is the use of these sensors to convert conventional spectrographs into spectropolarimeters.
This is worth considering carefully. Although calibration of instrumental polarization might be challenging in conventional grating spectrographs, it might be more tractable in grism spectrographs. 

\acknowledgments     

We thank the COLIBRÍ PI and deputy PI, Stéphane Basa and William Lee, for generously and patiently allowing us to commission {\TEQUILA} on COLIBRÍ.

We are grateful to Guillaume Molodij, Yannis Liodakis, and William Lee for comments which improved this presentation.

We thank Roger Blandford, Yannis Liodakis, A.N.~Ramaprakash, and Anthony Readhead for the stimulating early discussions that helped shape this project, and our SCEECS colleagues for their enthusiasm regarding the science applications of TEQUILA and organizing a discussion on jets and polarimetry at the annual meeting. 

We also thank our LAMA project collaborators, Dmitry Blinov, Elina Lindfors, Grigori Fedorets, Karri Koljonen, Heidi Korhonen, Yannis Liodakis, Kari Nilsson, and Klaas Wiersema, for illuminating discussions on the technical aspects of polarimetry with {\Polarsens} sensors and on the scientific opportunities enabled by them. 

AMW is grateful for the education in polarimetry he received decades ago from Art Code, Mário Magalhães, Ken Nordsieck, and Regina Schulte-Ladbeck at the University of Wisconsin---Madison. NG is grateful for the hands-on, experimental training in polarimetry gained through laboratory work at ELI Beamlines and in the Chemistry Department at UCSC.

NG sincerely thanks David Spergel for his continuous support and encouragement.
 
 We thank the staff of the Observatorio Astronómico Nacional at Sierra de San Pedro Mártir for their nightly devotion to operating COLIBRÍ. 
 
COLIBRÍ is an astronomical observatory developed and operated jointly by France (AMU, CNES, and CNRS) and Mexico (UNAM and SECIHTI). {\TEQUILA} was funded by the Simons Foundation (MP-SCMPS-00001470, N.G.),
CONACyT/SECIHTI 277901, and UNAM/DGAPA/PAPIIT projects, IN109224 and
IA101226.

\bibliography{report} 

@ARTICLE{1976JOSA...66..207N,
       author = {{Noll}, R.~J.},
        title = "{Zernike polynomials and atmospheric turbulence.}",
      journal = {Journal of the Optical Society of America (1917-1983)},
         year = 1976,
        month = mar,
       volume = {66},
        pages = {207-211},
          doi = {10.1364/JOSA.66.000207},
       adsurl = {https://ui.adsabs.harvard.edu/abs/1976JOSA...66..207N}
}

@ARTICLE{2020RMxAA..56..295P,
       author = {{Plauchu-Frayn}, Ilse and {Colorado}, Enrique and {Richer}, Michael G. and {Herrera-V{\'a}zquez}, Carlos},
        title = "{Thirteen Years of Weather Statistics at San Pedro Martir Observatory}",
      journal = {\rmxaa},
         year = 2020,
        month = oct,
       volume = {56},
        pages = {295-319},
          doi = {10.22201/ia.01851101p.2020.56.02.11},
archivePrefix = {arXiv},
       eprint = {2008.02402},
 primaryClass = {astro-ph.IM},
       adsurl = {https://ui.adsabs.harvard.edu/abs/2020RMxAA..56..295P}
}

@ARTICLE{2018PASA...35...12W,
       author = {{Wiersema}, K. and {Higgins}, A.~B. and {Covino}, S. and {Starling}, R.~L.~C.},
        title = "{Calibration of EFOSC2 Broadband Linear Imaging Polarimetry}",
      journal = {\pasa},
         year = 2018,
        month = mar,
       volume = {35},
          eid = {e012},
        pages = {e012},
          doi = {10.1017/pasa.2018.6},
archivePrefix = {arXiv},
       eprint = {1802.01890},
 primaryClass = {astro-ph.IM},
       adsurl = {https://ui.adsabs.harvard.edu/abs/2018PASA...35...12W}
}

@ARTICLE{2019MNRAS.482.5023H,
       author = {{Higgins}, A.~B. and {Wiersema}, K. and {Covino}, S. and {Starling}, R.~L.~C. and {Stevance}, H.~F. and {Wyrzykowski}, {\L}. and {Hodgkin}, S.~T. and {Maund}, J.~R. and {O'Brien}, P.~T. and {Tanvir}, N.~R.},
        title = "{SPLOT: a snapshot survey for polarized light in optical transients}",
      journal = {\mnras},
         year = 2019,
        month = feb,
       volume = {482},
       number = {4},
        pages = {5023-5040},
          doi = {10.1093/mnras/sty3029},
archivePrefix = {arXiv},
       eprint = {1811.01756},
 primaryClass = {astro-ph.HE},
       adsurl = {https://ui.adsabs.harvard.edu/abs/2019MNRAS.482.5023H}
}

@INPROCEEDINGS{2003SPIE.4843..456G,
       author = {{Giro}, Enrico and {Bonoli}, Carlotta and {Leone}, Franco and {Molinari}, Emilio and {Pernechele}, Claudio and {Zacchei}, Andrea},
        title = "{Polarization properties at the Nasmyth focus of the alt-azimuth TNG telescope}",
    booktitle = {Polarimetry in Astronomy},
         year = 2003,
       editor = {{Fineschi}, Silvano},
       series = {Society of Photo-Optical Instrumentation Engineers (SPIE) Conference Series},
       volume = {4843},
        month = feb,
        pages = {456-464},
          doi = {10.1117/12.458607},
       adsurl = {https://ui.adsabs.harvard.edu/abs/2003SPIE.4843..456G}
}

@software{Karpov2021,
       author = {{Karpov}, Sergey},
        title = "{STDPipe: Simple Transient Detection Pipeline}",
 howpublished = {Astrophysics Source Code Library, record ascl:2112.006},
         year = 2021,
        month = dec,
          eid = {ascl:2112.006},
archivePrefix = {ascl},
       eprint = {2112.006},
       adsurl = {https://ui.adsabs.harvard.edu/abs/2021ascl.soft12006K}
}

@INPROCEEDINGS{Watson2012,
       author = {{Watson}, Alan M. and {Richer}, Michael G. and {Bloom}, Joshua S. and {Butler}, Nathaniel R. and {Cese{\~n}a}, Urania and {Clark}, David and {Colorado}, Enrique and {C{\'o}rdova}, Antol{\'\i}n. and {Farah}, Alejandro and {Fox-Machado}, Lester and {Fox}, Ori D. and {Garc{\'\i}a}, Benjam{\'\i}n. and {Georgiev}, Leonid N. and {Gonz{\'a}lez}, J. Jes{\'u}s and {Guisa}, Gerardo and {Guti{\'e}rrez}, Leonel and {Herrera}, Joel and {Klein}, Christopher R. and {Kutyrev}, Alexander S. and {Lazo}, Francisco and {Lee}, William H. and {L{\'o}pez}, Eduardo and {Luna}, Esteban and {Mart{\'\i}nez}, Benjam{\'\i}n. and {Murillo}, Francisco and {Murillo}, Jos{\'e} Manuel and {N{\'u}{\~n}ez}, Juan Manuel and {Prochaska}, J. Xavier and {Ochoa}, Jos{\'e} Lu{\'\i}s. and {Quir{\'o}s}, Fernando and {Rapchun}, David A. and {Rom{\'a}n-Z{\'u}{\~n}iga}, Carlos and {Valyavin}, Gennady},
        title = "{Automation of the OAN/SPM 1.5-meter Johnson telescope for operations with RATIR}",
    booktitle = {Ground-based and Airborne Telescopes IV},
         year = 2012,
       editor = {{Stepp}, Larry M. and {Gilmozzi}, Roberto and {Hall}, Helen J.},
       series = {Society of Photo-Optical Instrumentation Engineers (SPIE) Conference Series},
       volume = {8444},
        month = sep,
          eid = {84445L},
        pages = {84445L},
          doi = {10.1117/12.926927},
       adsurl = {https://ui.adsabs.harvard.edu/abs/2012SPIE.8444E..5LW}
}

@ARTICLE{Basa2026,
       author = {{Basa}, S. and {Lee}, W.~H. and {Watson}, A.~M. and {Dolon}, F. and {Floriot}, J. and {Atteia}, J.-L. and {Dornic}, D. and {Lugo-Ibarra}, E.~E. and {Figueroa}, L. and {Langarica}, R. and {Valentin}, H. and {Ageron}, M. and {Agneray}, F. and {Alvarez Nunez}, L.~C. and {Angulo-Valdez}, C. and {Antier}, S. and {Auphan}, T. and {Baumann}, M. and {Bautista}, L. and {Becerra}, R.~L. and {Benahmed}, S. and {Benamar}, H. and {Blanpain}, C. and {Boulade}, O. and {Bounab}, Y. and {Boy}, J. and {Butler}, N.~R. and {Cadena Zepeda}, E.~O. and {Cuevas}, S. and {de Ugarte Postigo}, A. and {Delisle}, C. and {Devigny}, M. and {Ducoin}, J.~G. and {Fortin}, F. and {Fuentes-Fernandez}, J. and {Gaiti}, C. and {Garcia-Garcia}, L. and {Gallais}, P. and {Gill}, R. and {Globus}, N. and {Guisa}, G. and {Kajfasz}, E. and {Lafforgue}, D. and {Langlois}, A. and {Larrieu}, M. and {Landa}, J. and {Lecubin}, J. and {Lopez-Camara}, D. and {Lopez Angeles}, E. and {Lombardo}, S. and {Magnani}, F. and {Mandarakas}, N. and {Malgoyre}, A. and {Mathon}, R. and {Moreno Mendez}, E. and {Moreau}, C. and {Nouvel de la Fleche}, A. and {Ochoa}, J.~L. and {Ortiz}, L. and {Pedrayes-Lopez}, M.~H. and {Pereyra}, M. and {Provost}, L. and {Ramon}, P. and {Rakotondrainibe}, N.~A. and {Ronayette}, S. and {Ruiz Diaz-Soto}, J. and {Sanchez Alvarez}, F. and {Schneider}, B. and {Secroun}, A. and {Striebieg}, N. and {Tinoco}, S. and {Tourner-Sylvain}, M. and {Valenzuela}, F. and {Vincent}, D.},
        title = "{COLIBRI (SVOM/FM-GFT): Instrumentation and Performances on the SVOM Alerts}",
      journal = {arXiv e-prints},
         year = 2026,
        month = apr,
          eid = {arXiv:2604.24259},
        pages = {arXiv:2604.24259},
          doi = {10.48550/arXiv.2604.24259},
archivePrefix = {arXiv},
       eprint = {2604.24259},
 primaryClass = {astro-ph.IM},
       adsurl = {https://ui.adsabs.harvard.edu/abs/2026arXiv260424259B}
}

@ARTICLE{Fuentes2020,
       author = {{Fuentes-Fern{\'a}ndez}, J. and {Watson}, A.~M. and {Cuevas}, S. and {Basa}, S. and {Floriot}, J. and {Dolon}, F. and {Valentin}, H. and {Challita}, Z. and {Vola}, P.},
        title = "{Optical Design of COLIBR{\'I}: A Fast Follow-up Telescope for Transient Events}",
      journal = {Journal of Astronomical Instrumentation},
         year = 2020,
        month = jan,
       volume = {9},
       number = {1},
          eid = {2050001-280},
        pages = {2050001-280},
          doi = {10.1142/S2251171720500014},
       adsurl = {https://ui.adsabs.harvard.edu/abs/2020JAI.....950001F}
}

@software{bradley_2026_19636730,
  author       = {Bradley, Larry and
                  Sipőcz, Brigitta M. and
                  Robitaille, T. P. and
                  Tollerud, E. J. and
                  Vinícius, Zé and
                  Deil, Christoph and
                  Barbary, Kyle and
                  Wilson, Tom J. and
                  Busko, Ivo and
                  Donath, Axel and
                  Günther, Hans Moritz and
                  Cara, Mihai and
                  Lim, P. L. and
                  Meßlinger, Sebastian and
                  Conseil, Simon and
                  Droettboom, Michael and
                  Bostroem, K. Azalee and
                  Bray, E. M. and
                  Bratholm, Lars Andersen and
                  Burnett, Zach and
                  Jamieson, William and
                  Ginsburg, Adam and
                  Taranu, Dan and
                  Barentsen, Geert and
                  Craig, Matthew W. and
                  Morris, Brett M. and
                  Perrin, Marshall and
                  Rathi, Shivangee},
  title        = {Photutils},
  month        = apr,
  year         = 2026,
  publisher    = {Zenodo},
  version      = {3.0.0},
  doi          = {10.5281/zenodo.19636730},
  url          = {https://doi.org/10.5281/zenodo.19636730},
  swhid        = {swh:1:dir:5ea90432cd987336a26af201c1b50343b12e2daa
                   ;origin=https://doi.org/10.5281/zenodo.596036;visi
                   t=swh:1:snp:25dd84d95353ec3813baa8785c248017da7192
                   33;anchor=swh:1:rel:253cbde27398f9aea9d7368f386661
                   8c00fe16f9;path=astropy-photutils-3322558
                  },
}

@report{sony-data-sheet,
    author = {{Sony Semiconductor Solutions}},
    year = 2021,
    title = {{Sony} Polarization Image Sensor: {IMX250MZR/MYR}, {IMX264MZR/MYR}, {IMX253MZR/MYR}},
}

@report{qhy1253p-data-sheet,
    author = {QHYCCD},
    year = 2025,
    url = {https://www.qhyccd.com/scientific-camera-qhy1253-polarized-imx253/},
    title = {{QHY1253P} Data Sheet},
    publisher = {Light Speed Vision (Beijing) Co., Ltd.}
}

@INPROCEEDINGS{2024SPIE13096E..3DL,
       author = {{Langarica}, Rosal{\'\i}a. and {Watson}, Alan M. and {{\'A}lvarez N{\'u}{\~n}ez}, Luis Carlos and {Angeles}, Fernando and {Atteia}, Jean-Luc and {Basa}, St{\'e}phane and {Cuevas}, Salvador and {Dolon}, Fran{\c{c}}ois and {Farah}, Alejandro and {Dornic}, Damien and {Floriot}, Johan and {Fuentes-Fern{\'a}ndez}, Jorge and {Langlois}, Arthur and {Lee}, William H. and {Lombardo}, Simona and {Pereyra}, Margarita and {Ru{\'\i}z D{\'\i}az-Soto}, Jaime and {Ronayette}, Samuel and {Tinoco}, Silvio and {Valent{\'\i}n}, Herv{\'e}},
        title = "{The DDRAGO wide-field imager for the COLIBR{\'I} telescope}",
    booktitle = {Ground-based and Airborne Instrumentation for Astronomy X},
         year = 2024,
       editor = {{Bryant}, Julia J. and {Motohara}, Kentaro and {Vernet}, Jo{\"e}l. R.~D.},
       series = {Society of Photo-Optical Instrumentation Engineers (SPIE) Conference Series},
       volume = {13096},
        month = jul,
          eid = {130963D},
        pages = {130963D},
          doi = {10.1117/12.3020545},
       adsurl = {https://ui.adsabs.harvard.edu/abs/2024SPIE13096E..3DL}
}

@INPROCEEDINGS{2019ASSL..460...33B,
       author = {{Berdyugin}, Andrei and {Piirola}, Vilppu and {Poutanen}, Juri},
        title = "{Optical Polarimetry: Methods, Instruments and Calibration Techniques}",
    booktitle = {Astronomical Polarisation from the Infrared to Gamma Rays},
         year = 2019,
       editor = {{Mignani}, Roberto and {Shearer}, Andrew and {S{\l}owikowska}, Agnieszka and {Zane}, Silvia},
       series = {Astrophysics and Space Science Library},
       volume = {460},
        month = jan,
        pages = {33},
          doi = {10.1007/978-3-030-19715-5_3},
archivePrefix = {arXiv},
       eprint = {1908.10431},
 primaryClass = {astro-ph.IM},
       adsurl = {https://ui.adsabs.harvard.edu/abs/2019ASSL..460...33B}
}

@ARTICLE{2018ApOpt..57.4992F,
       author = {{Fei}, Huang and {Li}, Fan-Ming and {Chen}, Wei-Cong and {Zhang}, Rui and {Chen}, Chao-Shuai},
        title = "{Calibration method for division of focal plane polarimeters}",
      journal = {\ao},
         year = 2018,
        month = jun,
       volume = {57},
       number = {18},
        pages = {4992},
          doi = {10.1364/AO.57.004992},
       adsurl = {https://ui.adsabs.harvard.edu/abs/2018ApOpt..57.4992F}
}

@article{Malus1810,
  author  = {Étienne-Louis Malus},
  title   = {Théorie de la double réfraction de la lumière dans les substances cristallisées},
  journal = {Mémoires de l'Institut Impérial de France},
  volume  = {2},
  pages   = {143--158},
  year    = {1810}
}

@book{Arago1858,
  author  = {François Arago},
  editor  = {Jean-Augustin Barral},
  title   = {{Astronomie populaire}},
  volume  = {2},
  publisher = {Gide et J. Baudry},
  address = {Paris},
  year    = {1858},
  note    = {Livre XVII: ``Les com\`etes''}
}

@ARTICLE{2012ApOpt..51.5392Y,
       author = {{York}, Timothy and {Gruev}, Viktor},
        title = "{Characterization of a visible spectrum division-of-focal-plane polarimeter}",
      journal = {\ao},
         year = 2012,
        month = aug,
       volume = {51},
       number = {22},
        pages = {5392},
          doi = {10.1364/AO.51.005392},
       adsurl = {https://ui.adsabs.harvard.edu/abs/2012ApOpt..51.5392Y}
}

@INPROCEEDINGS{2011SPIE.8012E..0HY,
       author = {{York}, Timothy and {Gruev}, Viktor},
        title = "{Calibration method for division of focal plane polarimeters in the optical and near-infrared regime}",
    booktitle = {Infrared Technology and Applications XXXVII},
         year = 2011,
       editor = {{Andresen}, Bj{\o}rn F. and {Fulop}, Gabor F. and {Norton}, Paul R.},
       series = {Society of Photo-Optical Instrumentation Engineers (SPIE) Conference Series},
       volume = {8012},
        month = jun,
          eid = {80120H},
        pages = {80120H},
          doi = {10.1117/12.883950},
       adsurl = {https://ui.adsabs.harvard.edu/abs/2011SPIE.8012E..0HY}
}

@ARTICLE{2007PASP..119.1371T,
       author = {{Tinbergen}, J.},
        title = "{Accurate Optical Polarimetry on the Nasmyth Platform}",
      journal = {\pasp},
         year = 2007,
        month = dec,
       volume = {119},
       number = {862},
        pages = {1371-1384},
          doi = {10.1086/524225},
       adsurl = {https://ui.adsabs.harvard.edu/abs/2007PASP..119.1371T}
}

@ARTICLE{2023A&A...677A.144B,
       author = {{Blinov}, D. and {Maharana}, S. and {Bouzelou}, F. and {Casadio}, C. and {Gjerl{\o}w}, E. and {Jormanainen}, J. and {Kiehlmann}, S. and {Kypriotakis}, J.~A. and {Liodakis}, I. and {Mandarakas}, N. and {Markopoulioti}, L. and {Panopoulou}, G.~V. and {Pelgrims}, V. and {Pouliasi}, A. and {Romanopoulos}, S. and {Skalidis}, R. and {Anche}, R.~M. and {Angelakis}, E. and {Antoniadis}, J. and {Medhi}, B.~J. and {Hovatta}, T. and {Kus}, A. and {Kylafis}, N. and {Mahabal}, A. and {Myserlis}, I. and {Paleologou}, E. and {Papadakis}, I. and {Pavlidou}, V. and {Papamastorakis}, I. and {Pearson}, T.~J. and {Potter}, S.~B. and {Ramaprakash}, A.~N. and {Readhead}, A.~C.~S. and {Reig}, P. and {S{\l}owikowska}, A. and {Tassis}, K. and {Zensus}, J.~A.},
        title = "{The RoboPol sample of optical polarimetric standards}",
      journal = {\aap},
         year = 2023,
        month = sep,
       volume = {677},
          eid = {A144},
        pages = {A144},
          doi = {10.1051/0004-6361/202346778},
archivePrefix = {arXiv},
       eprint = {2307.06151},
 primaryClass = {astro-ph.IM},
       adsurl = {https://ui.adsabs.harvard.edu/abs/2023A&A...677A.144B}
}

@article{Nordin1999,
  author    = {Gregory P. Nordin and Jeffrey T. Meier and Panfilo C. Deguzman and Michael W. Jones},
  title     = {Micropolarizer array for infrared imaging polarimetry},
  journal   = {Journal of the Optical Society of America A},
  volume    = {16},
  number    = {5},
  pages     = {1168--1174},
  year      = {1999},
  publisher = {Optica Publishing Group},
  doi       = {10.1364/JOSAA.16.001168}
}

@inproceedings{Basa2022,
  author    = {St{\'e}phane Basa and William H. Lee and Fran{\c{c}}ois Dolon and
               A. M. Watson and J. Floriot and J. L. Atteia and others},
  title     = {{COLIBRI}, a wide-field 1.3 m robotic telescope dedicated to the transient sky},
  booktitle = {Ground-Based and Airborne Telescopes IX},
  editor    = {Heather K. Marshall and Jason Spyromilio and Tomonori Usuda},
  series    = {Proceedings of SPIE},
  volume    = {12182},
  pages     = {121821S},
  year      = {2022},
  publisher = {SPIE},
  doi       = {10.1117/12.2627139}
}

@inproceedings{Yamazaki2016,
  author    = {Tomohiro Yamazaki and Yasushi Maruyama and Yusuke Uesaka and others},
  title     = {Four-directional pixel-wise polarization {CMOS} image sensor using air-gap wire grid on 2.5-$\mu$m back-illuminated pixels},
  booktitle = {2016 IEEE International Electron Devices Meeting (IEDM)},
  year      = {2016},
  pages     = {8.7.1--8.7.4},
  publisher = {IEEE},
  doi       = {10.1109/IEDM.2016.7838378}
}

@ARTICLE{2013OExpr..2121039P,
       author = {{Powell}, S. Bear and {Gruev}, Viktor},
        title = "{Calibration methods for division-of-focal-plane polarimeters}",
      journal = {Optics Express},
         year = 2013,
        month = sep,
       volume = {21},
       number = {18},
        pages = {21039},
          doi = {10.1364/OE.21.021039},
       adsurl = {https://ui.adsabs.harvard.edu/abs/2013OExpr..2121039P}
}

@article{Bharathan2024,
title = {Simultaneous {X-ray} and optical polarization observations of the blazar {Mrk} 421},
author = {Bharathan, Athira M. and Stalin, C. S. and Sahayanathan, S. and Wani, Kiran and Mandal, Amit Kumar and Chatterjee, Rwitika and Joshi, Santosh and Pandey, Jeewan C. and Mathew, Blesson and Agrawal, Vivek K.},
journal = {arXiv preprint arXiv:2409.16745},
year = {2024},
archivePrefix = {arXiv},
eprint = {2409.16745},
primaryClass = {astro-ph.HE},
url = {https://arxiv.org/abs/2409.16745}
}

@article{Liodakis2019,
author = {Liodakis, Ioannis and Peirson, Abel L. and Romani, Roger W.},
title = {Prospects for Detecting {X-ray} Polarization in Blazar Jets},
journal = {The Astrophysical Journal},
volume = {880},
number = {1},
pages = {7},
year = {2019},
doi = {10.3847/1538-4357/ab2719},
eprint = {1906.01647},
archivePrefix = {arXiv},
primaryClass = {astro-ph.HE}
}

@inproceedings{Stockmans2024,
  author    = {Stockmans, Thijs and Scheinowitz, Naor and van der Linden, Erwoud and
               Malysheva, Irina and Strelow, Kira and Smit, Martijn and Snik, Frans},
  title     = {Sub-percent Characterization and Polarimetric Performance Analysis of Commercial Micro-polarizer Array Detectors},
  booktitle = {Polarization: Measurement, Analysis, and Remote Sensing XVI},
  year      = {2024},
  doi       = {10.1117/12.3013746}
}

@inproceedings{FORS2,
  author = {Appenzeller, I. et al.},
  title = {{FORS2} at the {VLT}},
  booktitle = {The Messenger},
  year = {1998}
}

@inproceedings{HOWPol,
  author = {Kawabata, K. S. et al.},
  title = {{HOWPol}: Hiroshima Optical and Near-Infrared Camera},
  booktitle = {SPIE},
  volume = {7014},
  year = {2008}
}

@article{RoboPol,
  author = {Ramaprakash, A. N. et al.},
  title = {The {RoboPol} instrument},
  journal = {MNRAS},
  volume = {485},
  pages = {2355--2366},
  year = {2019}
}

@article{YFPOL,
  author = {Xu, D. et al.},
  title = {{YFPOL}: A Linear Polarimeter of {Lijiang} 2.4 m Telescope},
  journal = {RAA},
  volume = {22},
  year = {2022}
}

@inproceedings{RINGO3,
  author = {Arnold, D. M. et al.},
  title = {{RINGO3}: A multi-colour fast response polarimeter},
  booktitle = {SPIE},
  volume = {8446},
  year = {2012}
}

@article{MOPTOP,
  author = {Jermak, H. et al.},
  title = {{MOPTOP}: a new polarimeter for the {Liverpool} {Telescope}},
  journal = {MNRAS},
  volume = {473},
  pages = {1269--1284},
  year = {2018}
}

@article{DIPOL1,
  author = {Piirola, V. et al.},
  title = {{DIPOL-1}: A double-image high-precision polarimeter},
  journal = {AJ},
  volume = {167},
  pages = {137},
  year = {2024}
}
\bibliographystyle{spiebib} 

\end{document}